\documentclass[12pt]{article}
\usepackage{epsfig,amssymb,psfrag,float,bm}
\usepackage[tbtags]{amsmath}

\usepackage{amsfonts,amsbsy,bm,euscript,mathrsfs}
\usepackage{amssymb,stmaryrd,faktor}
\usepackage{graphics}
\usepackage{tikz}
\usetikzlibrary{matrix,arrows}

\definecolor{hyperref}{RGB}{026,028,185}
\usepackage[bookmarks=true,colorlinks=true,linkcolor=hyperref,citecolor=hyperref,urlcolor=hyperref,bookmarksnumbered]{hyperref}

\numberwithin{equation}{section}

\def \be  {\begin{equation}}
\def \ee  {\end{equation}}
\def \ba  {\begin{eqnarray}}
\def \ea  {\end{eqnarray}}
\def\no{\nonumber}

\def\e{\epsilon}

\def\rmp{{\rm p}}

\newcommand{\arsinh}{\operatorname{arsinh}}
\newcommand{\csch}{\operatorname{csch}}
\newcommand{\sech}{\operatorname{sech}}

\newcommand{\IndA}{M}
\newcommand{\IndB}{N}
\newcommand{\IndC}{P}
\newcommand{\IndD}{Q}
\newcommand{\IndE}{R}
\newcommand{\IndF}{S}
\newcommand{\ms}{1}

\newcommand{\co}{\zeta}

\begin{document}

\thispagestyle{empty}

\vspace{ -3cm} \thispagestyle{empty} \vspace{-1cm}
\begin{flushright}
\footnotesize
HU-EP-13/17\\
\end{flushright}

\begingroup\centering
{\Large\bfseries\mathversion{bold}
Two-dimensional S-matrices from unitarity cuts
\par}
\vspace{7mm}

\begingroup
Lorenzo~Bianchi, Valentina~Forini, Ben~Hoare\\
\endgroup
\vspace{8mm}
\begingroup\small
\emph{Institut f\"ur Physik, 
Humboldt-Universit\"at
zu Berlin\\ Newtonstra{\ss}e 15, 12489 Berlin, Germany}\\
\endgroup

\vspace{0.3cm}
\begingroup\small
 {\tt $\{$lorenzo.bianchi, valentina.forini, ben.hoare$\}$@\,physik.hu-berlin.de}
\endgroup
\vspace{1.0cm}

\textbf{Abstract}\vspace{5mm}\par
\begin{minipage}{14.7cm}

Using unitarity methods, we compute, for several massive two-dimensional models,  the cut-constructible part of the one-loop $2\to 2$ scattering S-matrices from the tree-level amplitudes.
We apply our method to various integrable theories, finding evidence that for supersymmetric models the one-loop S-matrix is cut-constructible, while for models without supersymmetry (but with integrability) the missing rational terms are proportional to the tree-level S-matrix and therefore can be interpreted as a shift in the coupling.
Finally, applying our procedure to the world-sheet theory for the light-cone gauge-fixed AdS$_5\times S^5$ superstring we reproduce, at one-loop in the near-BMN expansion, the S-matrix known from integrability techniques.

\end{minipage}\par
\endgroup

\section{Introduction}
\label{sec:intro}

The remarkable efficiency of unitarity-based methods for the calculation of space-time scattering amplitudes in non-abelian gauge theories (see e.g.~\cite{Roib_Review}) motivates the application of 
similar techniques 
to perturbative regimes of other interesting models. 
This is certainly the case for the AdS$_5\times S^5$ superstring world-sheet theory. 
The S-matrix for the scattering of its excitations is known exactly, fixed by symmetry and integrability up to a phase \cite{Beisert:2005tm},
with the latter determined from a non-relativistic generalization of crossing symmetry as well as perturbative data 
both from the string and gauge theory sides -- see the review articles \cite{Arutyunov:2009ga,Beisert:2010jr} and references therein.
However, 
the perturbative study of scattering amplitudes computed from the path integral defined by the classical action is still of 
interest. This is true not only because such calculations serve as a test of the proposed exact quantum S-matrix, 
but also because they provide insights into 
the structure of the amplitudes and the manifestation of symmetries, and 
confirm the integrable setup~\cite{Klose:2006dd,Roiban:2006yc, Klose:2006zd,Klose:2007wq,Klose:2007rz}. 
The same is true for other integrable theories, including certain 2-d sigma-models~\cite{Zamolodchikov:1978xm,Ogievetsky:1987vv} 
and similar massive theories (see e.g.~\cite{Dorey:1996gd}), for which exact quantum S-matrices are known.
It is important to note that in the example of the light-cone gauge-fixed sigma-model 
for the AdS$_5\times S^5$ superstring standard perturbation theory 
has thus far not been a 
viable way -- due to regularization issues~\footnote{We thank Tristan McLoughlin and 
Radu Roiban for important discussions on this point.\label{foot1}} -- to evaluate the S-matrix beyond the leading order.

The aim of this work is to initiate the use of unitarity-based methods in the perturbative study 
of the S-matrix for \emph{massive two-dimensional} field theories~\footnote{Earlier attempts using unitarity to 
identify special structures in the amplitudes for the sine-Gordon model appeared in~\cite{Arefeva:1974bk}; 
we thank  Radu Roiban for pointing this out.}.
Confining ourselves to the use of \emph{standard} unitarity (where only two internal lines are placed on shell,  
subdividing a loop amplitude into two pieces~\footnote{In the case of \emph{generalized} unitarity~\cite{Britto:2004nc} 
(see e.g. \cite{Bern:2011qt})  the loop amplitude is subdivided into more than two pieces, which corresponds to placing multiple internal lines on-shell.}) and working in a \emph{fixed} number of dimensions $d=2$, we propose a formula for constructing the one-loop $2\to 2$ scattering amplitude 
directly from the corresponding on-shell tree-level amplitudes.  

As might be expected, the  two-dimensional case is much simpler than its four-dimensional counterpart.  Our proposals,  equations \eqref{eqn:final} and \eqref{eqn:final_ferm} below, take a remarkably compact form. 
This is a consequence of the massive 2-d kinematics, which imply that the cut loop momenta are frozen to specific values.
Therefore, the integral  degenerates to a sum over discrete solutions of the on-shell conditions. 
This is reminiscent of the  framework  of generalized unitarity in the four-dimensional case when quadruple  cuts (maximal cuts~\cite{Britto:2004nc}) are used. There, the quadruple-cut integral is completely localized by the four delta-functions of the cut propagators, and it reduces to a product of four tree-level amplitudes. 

It is important to note that our procedure is inherently finite. All the theories we 
consider are 
either UV-finite or renormalizable. 
In the latter case, our procedure implicitly chooses a particular regularization scheme. Hence the result we will compare to is the renormalized four-point amplitude, up to scheme ambiguities.  
A related point is that the expressions \eqref{eqn:final} and \eqref{eqn:final_ferm}, obtained via the implementation of unitary cuts in fixed dimension, are not necessarily expected to give the full answer, possibly missing rational terms 
(terms  with only a rational dependence on momentum invariants) that can arise from the $\epsilon$-expansion in $d=2-2\epsilon$ dimensions~\cite{vanNeerven:1985xr}. A thorough discussion of these issues is contained in Section~\ref{sec:general}.
 
The main part of this paper focusses on testing the validity of our procedure for several  models of interest, and therefore exploring 
the \emph{cut-constructibility} (via standard unitarity methods) of S-matrices in two dimensions. 
For bosonic theories with \emph{integrability}, the result obtained is rather intriguing -- we find close agreement with perturbation theory,
the only price to pay being a finite shift in the coupling.
Incidentally, a hint of a connection between unitarity and integrability can be seen, for example, in the sine-Gordon model -- the only non-vanishing contribution to the connected part of the $3 \to 3$  tree-level  amplitude is obtained by setting the intermediate particles on-shell~\cite{Dorey:1996gd}.
These results seem to suggest a relationship between quantization preserving integrability and unitarity techniques which would be interesting to investigate further.

Possibly more expected is that one-loop amplitudes in \emph{integrable, supersymmetric} theories appear to be cut-constructible via standard unitarity. This is analogous to the cut-constructibility of one-loop amplitudes in massless, supersymmetric theories in four dimensions~\cite{Bern:1994zx,Bern:1994cg}. 
We also apply our procedure to the AdS$_5 \times S^5$ light-cone gauge-fixed world-sheet superstring finding agreement, at one-loop in the near-BMN expansion, with the S-matrix known from integrability techniques.

\bigskip

The paper proceeds as follows. In Section \ref{sec:general} we set out the general formalism, and obtain the expressions for the cut-constructible part of the one-loop $2\to 2$ S-matrix. We then apply this result to various relativistic integrable models in Section \ref{sec:relativistic} and to string theory on AdS$_5\times S^5$ in Section \ref{sec:strings}.  Concluding remarks are given in Section \ref{sec:conclusions}.

 
\section{Two-particle one-loop S-matrix from unitarity cuts}
\label{sec:general}

In this section we derive a candidate expression for the one-loop two-particle S-matrix in terms of the tree-level one, as follows from the standard application of cutting techniques (see for example ~\cite{Dixon:1996wi}) to the two-dimensional case. While due to the expected absence of rational terms this formula can never be completely general, we investigate its validity in later sections on various examples.\\
The two-body scattering process of a field theory invariant under space and time translations
\be\label{eqn:4ptampl}
\langle\Phi^\IndC(p_3)\Phi^\IndD(p_4)\,|\mathbb{S}|\,\Phi_\IndA(p_1)\Phi_\IndB(p_2)\rangle=\mathcal{A}_{\IndA\IndB}^{\IndC\IndD}(p_1,p_2,p_3,p_4)
\ee  
is described via the four-point amplitude
\be\label{eqn:ampcons}
\mathcal{A}_{\IndA\IndB}^{\IndC\IndD}(p_1,p_2,p_3,p_4)=(2\pi)^2 \delta^{(d)}(p_1+p_2-p_3-p_4)\, \widetilde{\mathcal{A}}_{\IndA\IndB}^{\IndC\IndD}(p_1,p_2,p_3,p_4)~.
\ee
In (\ref{eqn:4ptampl}) $\mathbb{S}$ is the scattering operator, the fields $\Phi$ carry flavor indices to account for different kinds of particles in the model and $p_i$ are their on-shell momenta. In this paper we will restrict to the case where all the particles  have equal non-vanishing mass, which we set to unity. 
In the two-dimensional case the set of initial momenta is preserved under collision. This translates into the following identity for the energy-momentum conservation $\delta$-function of (\ref{eqn:ampcons}) 
\be\label{delta2d}
\delta^{(2)}(p_1+p_2-p_3-p_4)=J(p_1,p_2)\,\big(\delta(\text{p}_1 - \text{p}_3)\delta(\text{p}_2 - \text{p}_4) +\delta(\text{p}_1 - \text{p}_4)\delta(\text{p}_2 - \text{p}_3) \big) \ .
\ee
Above, $\rmp$ is the spatial momentum and the Jacobian $J(p_1,p_2)=1/(\partial \e_{\rmp_1}/\partial \rmp_1-\partial\e_{\rmp_2}/\partial \rmp_2)$ depends on the dispersion relation $\e_\rmp$  (the on-shell energy associated to $\rmp$) for the theory at hand. Spatial momenta are assumed to be ordered $\rmp_1>\rmp_2$.  Substituting (\ref{delta2d}) in \eqref{eqn:ampcons}  we can consider the amplitudes associated to the first product of $\delta$-functions $\delta(\text{p}_1 - \text{p}_3)\delta(\text{p}_2 - \text{p}_4)$  without loss of generality.
The S-matrix elements relevant for the description of the $2\to2$ scattering in the two-dimensional case are then defined as \cite{Eden:1966}
\be\label{AandS}
S_{\IndA\IndB}^{\IndC\IndD}(p_1,p_2)\equiv \frac{J(p_1,p_2)}{4\e_1 \e_2} \widetilde{\mathcal{A}}_{\IndA\IndB}^{\IndC\IndD}(p_1,p_2,p_1,p_2) ~,
\ee
where the denominator is required to make contact with the standard definition of the S-matrix in two dimensions.
In applying the unitarity method to the one-loop four point amplitude (\ref{eqn:ampcons}) one follows the standard route of considering two-particle cuts, obtained by putting two intermediate lines on-shell. 

\begin{figure}[ht]
\begin{center}
\begin{tikzpicture}[line width=2pt,scale=1.5]
\draw[-] (-5,-3) -- (-4,-4);
\draw[-] (-5,-5) -- (-4,-4);
\draw[-] (-2,-4) -- (-1,-3);
\draw[-] (-2,-4) -- (-1,-5);
\draw    (-3,-4) circle (1cm);
\draw[|-|,dashed,red!70,line width=1pt] (-3,-2.5) -- (-3,-5.5);
\draw[->] (-4.7,-3.0) -- (-4.3,-3.4);
\node at (-4.7,-2.8) {$p_1$};
\draw[->] (-4.7,-5.0) -- (-4.3,-4.6);
\node at (-4.7,-5.2) {$p_2$};
\draw[<-] (-1.3,-3.0) -- (-1.7,-3.4);
\node at (-1.3,-2.8) {$p_4$};
\draw[<-] (-1.3,-5.0) -- (-1.7,-4.6);
\node at (-1.3,-5.2) {$p_3$};
\draw[<-] (-3.1,-3.2) -- (-3.5,-3.35);
\node at (-3.2,-3.5) {$l_1$};
\draw[->] (-2.9,-4.8) -- (-2.5,-4.65);
\node at (-2.8,-4.5) {$l_2$};
\node at (-3.4,-2.85) {$\IndE$};
\node at (-2.6,-5.15) {$\IndF$};
\node at (-5.2,-3) {$\IndA$};
\node at (-5.2,-5) {$\IndB$};
\node at (-0.8,-5) {$\IndC$};
\node at (-0.8,-3) {$\IndD$};
\node [circle,draw=black!100,fill=black!5,thick,opacity=.95] at (-4,-4) {\footnotesize$\mathcal{A}^{(0)}$\normalsize};
\node [circle,draw=black!100,fill=black!5,thick,opacity=.95] at (-2,-4) {\footnotesize$\mathcal{A}^{(0)}$\normalsize};
\end{tikzpicture}
\\
\begin{tikzpicture}[line width=2pt,scale=1.5,rotate=90]
\draw[-] (-5,-3) -- (-4,-4);
\draw[-] (-5,-5) -- (-4,-4);
\draw[-] (-2,-4) -- (-1,-3);
\draw[-] (-2,-4) -- (-1,-5);
\draw    (-3,-4) circle (1cm);
\draw[|-|,dashed,red!70,line width=1pt] (-3,-2.5) -- (-3,-5.5);
\draw[->] (-4.7,-3.0) -- (-4.3,-3.4);
\node at (-4.7,-2.8) {$p_2$};
\draw[<-] (-4.7,-5.0) -- (-4.3,-4.6);
\node at (-4.7,-5.2) {$p_4$};
\draw[->] (-1.3,-3.0) -- (-1.7,-3.4);
\node at (-1.3,-2.8) {$p_1$};
\draw[<-] (-1.3,-5.0) -- (-1.7,-4.6);
\node at (-1.3,-5.2) {$p_3$};
\draw[<-] (-3.1,-3.2) -- (-3.5,-3.35);
\node at (-3.2,-3.5) {$l_1$};
\draw[<-] (-2.9,-4.8) -- (-2.5,-4.65);
\node at (-2.8,-4.5) {$l_2$};
\node at (-3.4,-2.85) {$\IndE$};
\node at (-2.6,-5.15) {$\IndF$};
\node at (-5.2,-3) {$\IndB$};
\node at (-5.2,-5) {$\IndD$};
\node at (-0.8,-5) {$\IndC$};
\node at (-0.8,-3) {$\IndA$};
\node [circle,draw=black!100,fill=black!5,thick,opacity=.95] at (-4,-4) {\footnotesize$\mathcal{A}^{(0)}$\normalsize};
\node [circle,draw=black!100,fill=black!5,thick,opacity=.95] at (-2,-4) {\footnotesize$\mathcal{A}^{(0)}$\normalsize};
\end{tikzpicture}
\hspace{10pt}
\begin{tikzpicture}[line width=2pt,scale=1.5,rotate=90]
\draw[-] (-5,-3) -- (-4,-4);
\draw[-] (-5,-5) -- (-4,-4);
\draw[-] (-2,-4) -- (-1,-3);
\draw[-] (-2,-4) -- (-1,-5);
\draw    (-3,-4) circle (1cm);
\draw[|-|,dashed,red!70,line width=1pt] (-3,-2.5) -- (-3,-5.5);
\draw[->] (-4.7,-3.0) -- (-4.3,-3.4);
\node at (-4.7,-2.8) {$p_2$};
\draw[<-] (-4.7,-5.0) -- (-4.3,-4.6);
\node at (-4.7,-5.2) {$p_3$};
\draw[->] (-1.3,-3.0) -- (-1.7,-3.4);
\node at (-1.3,-2.8) {$p_1$};
\draw[<-] (-1.3,-5.0) -- (-1.7,-4.6);
\node at (-1.3,-5.2) {$p_4$};
\draw[<-] (-3.1,-3.2) -- (-3.5,-3.35);
\node at (-3.2,-3.5) {$l_1$};
\draw[<-] (-2.9,-4.8) -- (-2.5,-4.65);
\node at (-2.8,-4.5) {$l_2$};
\node at (-3.4,-2.85) {$\IndE$};
\node at (-2.6,-5.15) {$\IndF$};
\node at (-5.2,-3) {$\IndB$};
\node at (-5.2,-5) {$\IndC$};
\node at (-0.8,-5) {$\IndD$};
\node at (-0.8,-3) {$\IndA$};
\node [circle,draw=black!100,fill=black!5,thick,opacity=.95] at (-4,-4) {\footnotesize$\mathcal{A}^{(0)}$\normalsize};
\node [circle,draw=black!100,fill=black!5,thick,opacity=.95] at (-2,-4) {\footnotesize$\mathcal{A}^{(0)}$\normalsize};
\end{tikzpicture}
\caption{Diagrams representing s-, t- and u-channel cuts contributing to the four-point one-loop amplitude.}
\label{stu}\nonumber
\end{center}
\end{figure}
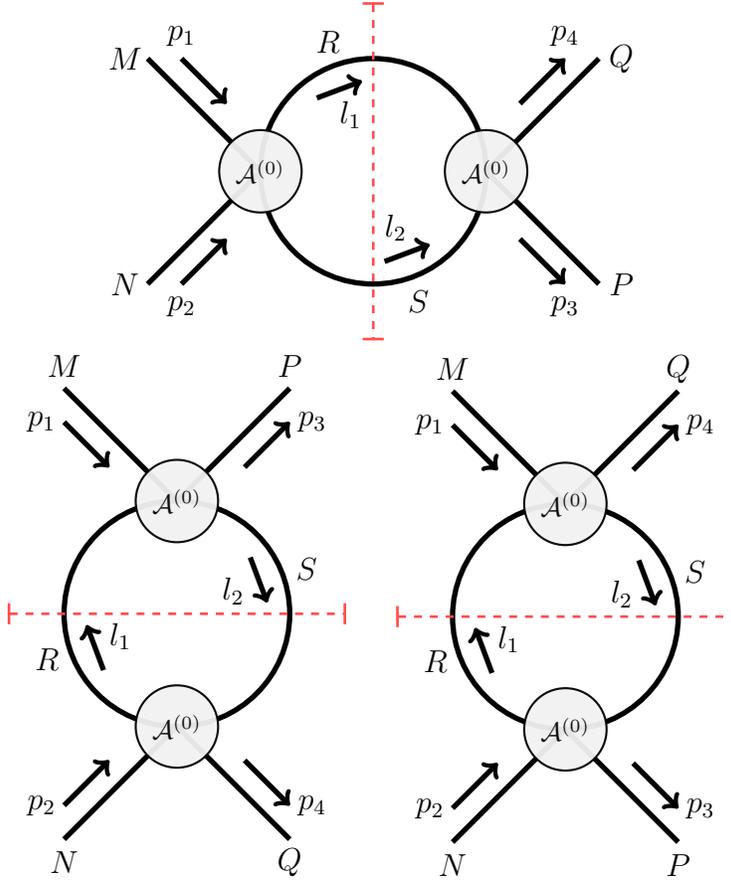

In general there will be contributions from tadpole and bubble graphs to the one-loop four point amplitude. The former have no physical two-particle cuts and do not contribute an imaginary part to the amplitude. For this reason they should not be considered in our procedure, which is based on standard unitarity rules (derived from the optical theorem) \cite{Bern:1994zx,Bern:1997nh}. Therefore the one-loop result following from unitarity techniques will receive contributions from the $s$- $t$- and $u$-  channel cuts illustrated in Fig.~\ref{stu}. Explicitly,
the imaginary part of the amplitude is given by the sum of the following three contributions
 \ba\no
\mathcal{A}^{(1)}{}^{\IndC\IndD}_{\IndA\IndB}(p_1,p_2,p_3,p_4)|_{s-cut}=\frac12\int\frac{d^2 l_1}{(2\pi)^2}\int\frac{d^2 l_2}{(2\pi)^2}\ i\pi\delta^+({l_1}^2-\ms)\ i\pi\delta^+(l_2^2-\ms)\\\label{eqn:sch1}
\times\,\mathcal{A}^{(0)}{}_{\IndA\IndB}^{\IndE\IndF}({p_1,p_2,l_1,l_2})\mathcal{A}^{(0)}{}_{\IndF\IndE}^{\IndC\IndD} ({l_2,l_1,p_3,p_4})
\\\no
\mathcal{A}^{(1)}{}^{\IndC\IndD}_{\IndA\IndB}(p_1,p_2,p_3,p_4)|_{t-cut}=\frac12\int\frac{d^2 l_1}{(2\pi)^2}\int\frac{d^2 l_2}{(2\pi)^2}\ i\pi\delta^+({l_1}^2-\ms)\ i\pi\delta^+({l_2}^2-\ms)\\\label{eqn:tch1}
\times\,\mathcal{A}^{(0)}{}_{\IndA\IndE}^{\IndF\IndC}({p_1,l_1,l_2,p_3})\mathcal{A}^{(0)}{}_{\IndF\IndB}^{\IndE\IndD}({l_2,p_2,l_1,p_4})\\\no
\mathcal{A}^{(1)}{}^{\IndC\IndD}_{\IndA\IndB}(p_1,p_2,p_3,p_4)|_{u-cut}=\frac12\int\frac{d^2 l_1}{(2\pi)^2}\int\frac{d^2 l_2}{(2\pi)^2}\ i\pi\delta^+({l_1}^2-\ms)\ i\pi\delta^+({l_2}^2-\ms)\\\label{eqn:uch1}
\times\,\mathcal{A}^{(0)}{}_{\IndA\IndE}^{\IndF\IndD}({p_1,l_1,l_2,p_4})\mathcal{A}^{(0)}{}_{\IndF\IndB}^{\IndE\IndC}({l_2,p_2,l_1,p_3})
\ea
where $\mathcal{A}^{(0)}$ are tree-level amplitudes and a sum over the complete set of intermediate states $\IndE,\IndF$ (all allowed particles for the cut lines) is understood. The on-shell propagator is given in terms of $\delta^+(k^2 - 1) = \theta(k^0) \delta(k^2-1)$ and we have included a symmetry factor of $\tfrac12$.

To proceed, in each case we use (\ref{eqn:ampcons}) and the two-momentum conservation at the vertex involving the momentum $p_1$ to integrate over $l_2$
\ba
 &&\!\!\!\!\!\!\!\!\!\!\!\!\!\!\!\!\!\!\!\!\!\!\!
 \mathcal{\widetilde{A}}^{(1)}{}^{\IndC\IndD}_{\IndA\IndB}(p_1,p_2,p_3,p_4)|_{s-cut}=\frac12\int\frac{d^2 l_1}{(2\pi)^2}\,i\pi\delta^+({l_1}^2-\ms)\, i\pi\delta^+(({l_1}-{p_1}-{p_2})^2-\ms)\nonumber\\
&&\qquad\times \,\,\widetilde{\mathcal{A}}^{(0)}{}_{\IndA\IndB}^{\IndE\IndF}({p_1,p_2,l_1,-l_1+p_1+p_2}) \,\widetilde{\mathcal{A}}^{(0)}{}_{\IndF\IndE}^{\IndC\IndD}({-l_1+p_1+p_2,l_1,p_3,p_4})\,,\label{2_8} \\
 &&\!\!\!\!\!\!\!\!\!\!\!\!\!\!\!\!\!\!\!\!\!\!\!
 \mathcal{\widetilde{A}}^{(1)}{}^{\IndC\IndD}_{\IndA\IndB}(p_1,p_2,p_3,p_4)|_{t-cut}=\frac12\int\frac{d^2 l_1}{(2\pi)^2}\,i\pi\delta^+({l_1}^2-\ms)\,i\pi\delta^+(({l_1}+{p_1}-{p_3})^2-\ms)   \nonumber\\
&&\qquad\times\,\, \widetilde{\mathcal{A}}^{(0)}{}_{\IndA\IndE}^{\IndF\IndC}({p_1,l_1,l_1+p_1-p_3,p_3})\,  \widetilde{\mathcal{A}}^{(0)}{}_{\IndF\IndB}^{\IndE\IndD}({l_1+p_1-p_3,p_2,l_1,p_4})\, , \label{2_9} \\
 &&\!\!\!\!\!\!\!\!\!\!\!\!\!\!\!\!\!\!\!\!\!\!\!
 \mathcal{\widetilde{A}}^{(1)}{}^{\IndC\IndD}_{\IndA\IndB}(p_1,p_2,p_3,p_4)|_{u-cut}=
\frac12\int\frac{d^2 l_1}{(2\pi)^2}\,i\pi\delta^+({l_1}^2-\ms)\, i\pi\delta^+(({l_1}+{p_1}-{p_4})^2-\ms)\nonumber\\
&&\qquad\times\,\, \widetilde{\mathcal{A}}^{(0)}{}_{\IndA\IndE}^{\IndF\IndD}({p_1,l_1,l_1+p_1-p_4,p_4})\,\widetilde{\mathcal{A}}^{(0)}{}_{\IndF\IndB}^{\IndE\IndC}({l_1+p_1-p_4,p_2,l_1,p_3})\,.\label{2_10}
\ea
In each of these integrals the set of zeroes of the $\delta$-functions are discrete. This allows us to pull out the tree-level amplitudes with the loop-momenta evaluated at those zeroes, leaving scalar bubbles~\footnote{Note that if one first uses the $\delta$-function identity \eqref{delta2d} to fix, for example, $p_1 = p_3$ and $p_2 = p_4$ the $t$-cut integral is ill-defined. Furthermore, the procedure of fixing $l_1 = p_3$ no longer follows. Therefore, to avoid this ambiguity we follow the prescription that we should only impose the $\delta$-function identity \eqref{delta2d} at the end. In some sense this is natural as, in general dimensions, QFT amplitudes have the form \eqref{eqn:ampcons}, while the $\delta$-function identity \eqref{delta2d} is specific to two dimensions.}. Following standard unitarity computations \cite{Bern:1994zx}, we apply the following replacement in the imaginary part of the amplitude \eqref{2_8}--\eqref{2_10} to the internal on-shell propagators:
$i\pi \delta^+(l^2-1) \longrightarrow \tfrac{1}{l^2-1}$.
This allows us to rebuild, from its imaginary part, the cut-constructible piece of the amplitude
\ba\nonumber
 \mathcal{\widetilde{A}}^{(1)}{}^{\IndC\IndD}_{\IndA\IndB}(p_1,p_2,p_3,p_4)&=&\frac{I(p_1+p_2)}{4}
\Big[ \widetilde{\mathcal{A}}^{(0)}{}_{\IndA\IndB}^{\IndE\IndF}({p_1,p_2,p_1,p_2}) \widetilde{\mathcal{A}}^{(0)}{}_{\IndF\IndE}^{\IndC\IndD}({p_2,p_1,p_3,p_4})\\ \no
&&\qquad\qquad~+\,  \widetilde{\mathcal{A}}{}^{(0)}{}_{\IndA\IndB}^{\IndE\IndF}({p_1,p_2,p_2,p_1}) \widetilde{\mathcal{A}}^{(0)}{}_{\IndF\IndE}^{\IndC\IndD}({p_1,p_2,p_3,p_4}) \Big] \\\no
& +&\frac{I(p_1-p_3)}{2}   \widetilde{\mathcal{A}}{}^{(0)}{}_{\IndA\IndE}^{\IndF\IndC}({p_1,p_3,p_1,p_3}) \widetilde{\mathcal{A}}^{(0)}{}_{\IndF\IndB}^{\IndE\IndD}({p_1,p_2,p_3,p_4})\\ 
& + &\frac{ I(p_1-p_4)}{2}   \widetilde{\mathcal{A}}{}^{(0)}{}_{\IndA\IndE}^{\IndF\IndD}({p_1,p_4,p_1,p_4}) \widetilde{\mathcal{A}}^{(0)}{}_{\IndF\IndB}^{\IndE\IndC}({p_1,p_2,p_4,p_3})\label{2_11}
\ea
where we have introduced the bubble integral
\be
I(p)=\int \frac{d^2 q}{(2\pi)^2} \frac{1}{(q^2-\ms+i\e) ((q-p)^2-\ms+i\e)}~
\ee 
The structure of \eqref{2_11} shows the difference between the $s$-channel, for which there are two solutions of the $\delta$-function constraints in \eqref{2_8} (for positive energies), and the $t$- and $u$-channels, for which there is only one. 

Choosing  $p_3=p_1$, $p_4=p_2$, 
which corresponds to  considering the amplitudes associated to the first product of $\delta$-functions $\delta(\text{p}_1 - \text{p}_3)\delta(\text{p}_2 - \text{p}_4)$ (see comment above \eqref{AandS}), 
 it then follows that a candidate expression for the
one-loop S-matrix elements is given by the following simple sum of products of two tree-level amplitudes weighted by scalar bubble integrals.
\begin{align}
 {S^{(1)}}{}_{\IndA\IndB}^{\IndC\IndD}(p_1,p_2)=\frac{1}{8 (\e_2\,\rmp_1-\e_1\,\rmp_2)}\,\Big[&{\tilde S^{(0)}}{}_{\IndA\IndB}^{\IndE\IndF}(p_1,p_2){\tilde S^{(0)}}{}_{\IndE\IndF}^{\IndC\IndD}(p_1,p_2)I(p_1+p_2)\nonumber\\ \vphantom{\frac{1}{4 (\e_2\,\rmp_1-\e_1\,\rmp_2)}} 
 +& {\tilde S^{(0)}}{}_{\IndA\IndE}^{\IndF\IndC}(p_1,p_1){\tilde S^{(0)}}{}_{\IndF\IndB}^{\IndE\IndD}(p_1,p_2)I(0) \nonumber\\
 \vphantom{\frac{1}{4 (\e_2\,\rmp_1-\e_1\,\rmp_2)}}
 +&{\tilde S^{(0)}}{}_{\IndA\IndE}^{\IndF\IndD}(p_1,p_2){\tilde S^{(0)}}{}_{\IndF\IndB}^{\IndC\IndE}(p_1,p_2)I(p_1-p_2)\,\Big]
 \label{eqn:final}~,
\end{align}
where $\tilde S^{(0)}(p_1,p_2)=4 (\e_2\,\rmp_1-\e_1\,\rmp_2) S^{(0)}(p_1,p_2)$.
The denominator on the right-hand side comes from the Jacobian $J(p_1,p_2)$ assuming a standard relativistic dispersion relation (for the theories we consider, at one-loop this is indeed the case), and 
the one-loop integrals  read explicitly 
\begin{align}
 I(p_1+p_2)&=\frac{i\pi-\arsinh(\e_2\,\rmp_1-\e_1\,\rmp_2)}{4\pi i\,(\e_2\,\rmp_1-\e_1\,\rmp_2)}\,,\label{bubblep}\\
 I(0)&=\frac{1}{4\pi i}\,,\\
 I(p_1-p_2)&=\frac{\arsinh(\e_2\,\rmp_1-\e_1\,\rmp_2)}{4\pi i\,(\e_2\,\rmp_1-\e_1\,\rmp_2)}~.\label{bubblem}
\end{align}
For theories including fermionic fields, the above derivation holds up to signs. To be precise the sign prescription is given as follows: 
\begin{align}\no
 {S^{(1)}}{}_{\IndA\IndB}^{\IndC\IndD}(p_1,p_2)=&\frac{1}{8 (\e_2\,\rmp_1-\e_1\,\rmp_2)}\,\Big[{\tilde S^{(0)}}{}_{\IndA\IndB}^{\IndE\IndF}(p_1,p_2){\tilde S^{(0)}}{}_{\IndE\IndF}^{\IndC\IndD}(p_1,p_2)I(p_1+p_2)\nonumber\\\label{eqn:final_ferm}
  \vphantom{\frac{1}{4 (\e_2\,\rmp_1-\e_1\,\rmp_2)}}
 +&  (-1)^{[\IndC][\IndF]+[\IndE][\IndF]}\,{\tilde S^{(0)}}{}_{\IndA\IndE}^{\IndF\IndC}(p_1,p_1){\tilde S^{(0)}}{}_{\IndF\IndB}^{\IndE\IndD}(p_1,p_2)I(0) \\ \no
 \vphantom{\frac{1}{4 (\e_2\,\rmp_1-\e_1\,\rmp_2)}}
 +&(-1)^{[\IndC][\IndE]+[\IndD][\IndF]+[\IndE][\IndF]+[\IndC][\IndD]}{\tilde S^{(0)}}{}_{\IndA\IndE}^{\IndF\IndD}(p_1,p_2){\tilde S^{(0)}}{}_{\IndF\IndB}^{\IndC\IndE}(p_1,p_2)I(p_1-p_2)\,\Big]~,
\end{align}
where $[\IndA] = 0$ for a boson and $1$ for a fermion. 

The expressions \eqref{eqn:final} and \eqref{eqn:final_ferm} above, as is clear from the second term, are not invariant under the   interchange of $p_1$ and $p_2$ along with the corresponding flavor indices. Furthermore, if we choose the alternative solution of the conservation $\delta$-function in (\ref{eqn:tch1}), namely $\ell_2=\ell_1+p_4-p_2$, the coefficient of $I(0)$ in (\ref{eqn:final}) would read
\be
{\tilde S^{(0)}}{}_{\IndA\IndE}^{\IndC\IndF}(p_1,p_2){\tilde S^{(0)}}{}_{\IndF\IndB}^{\IndD\IndE}(p_2,p_2)~,
\ee
where, as before, for theories including fermionic fields there is an additional sign given by $(-1)^{[\IndD][\IndE]+[\IndE][\IndF]}$.
Therefore, consistency between the two expressions requires the following condition on the tree-level S-matrix 
\be\label{consistency}
{\tilde S^{(0)}}{}_{\IndA\IndE}^{\IndF\IndC}(p_1,p_1)\,{\tilde S^{(0)}}{}^{\IndE\IndD}_{\IndF\IndB}(p_1,p_2)\,=\,{\tilde S^{(0)}}{}^{\IndC\IndF}_{\IndA\IndE}(p_1,p_2)\,{\tilde S^{(0)}}{}^{\IndD\IndE}_{\IndF\IndB}(p_2,p_2)~.
\ee
In the case where there are also fermionic fields the consistency condition is generalized to
\be\label{consistency_ferm}
(-1)^{[\IndC][\IndF]+[\IndE][\IndF]}\,{\tilde S^{(0)}}{}_{\IndA\IndE}^{\IndF\IndC}(p_1,p_1)\,{\tilde S^{(0)}}{}^{\IndE\IndD}_{\IndF\IndB}(p_1,p_2)\,=
(-1)^{[\IndD][\IndE]+[\IndE][\IndF]}
\,{\tilde S^{(0)}}{}^{\IndC\IndF}_{\IndA\IndE}(p_1,p_2)\,{\tilde S^{(0)}}{}^{\IndD\IndE}_{\IndF\IndB}(p_2,p_2)~.
\ee
If \eqref{consistency}/\eqref{consistency_ferm} is satisfied, then the formula (\ref{eqn:final})/\eqref{eqn:final_ferm} is free from ambiguities.
We have checked this for the tree-level S-matrices of all the field theory models treated below.

Finally, let us comment on the issue of renormalization. It is clear that the results, \eqref{eqn:final} and \eqref{eqn:final_ferm}, following from our procedure are finite quantities as they only involve the scalar bubble integral in two dimensions. This follows from the discreteness of the set of zeroes of the arguments of the $\delta$-functions in \eqref{2_8}-\eqref{2_10},   which allows one to  pull out the tree-level amplitudes from the loop integral and was  used to derive \eqref{2_11}. Consequently, no additional regularization is required and the result can be compared directly with the $2 \to 2$ particle S-matrix (following from the finite or renormalized four-point amplitude) found using standard perturbation theory.

Of course, this need not be the case for the original bubble integrals before cutting -- due to factors of loop-momentum in the numerators. These divergences, along with those coming from tadpole graphs, which we did not consider, should be taken into account for the renormalization of the theory. In this paper we do not investigate this issue, however all the theories we consider in Section \ref{sec:relativistic} and \ref{sec:strings} are either UV-finite or renormalizable. Furthermore, whether the cut-constructible contribution \eqref{eqn:final}/\eqref{eqn:final_ferm} to the S-matrix gives the full result is a priori unclear as rational terms following from non-trivial cancellations in the regularization procedure may be missing.

In the following sections we will make use of (\ref{eqn:final}) and \eqref{eqn:final_ferm} to construct the one-loop S-matrix from its tree-level form for some relativistic models (Section \ref{sec:relativistic}) and for the non-relativistic string world-sheet field theory on AdS$_5\times S^5$ (Section \ref{sec:strings}). For all these theories we will compare with known results (either from perturbation theory or integrability techniques) and analyze the effectiveness of the method of unitarity cuts in two dimensions.

\section{Relativistic models}
\label{sec:relativistic}

To explore the validity of the formula we will initially focus on some relativistic models. In relativistic theories it is natural to write the on-shell momenta in terms of rapidities
\begin{equation}
{\text p}_i = \sinh \vartheta_i \ , \qquad \epsilon_{{\text p}_i} = \cosh \vartheta_i \ ,
\end{equation}
where the mass has been set to unity. Lorentz invariance then implies that the $2\to 2$ scattering S-matrix should depend solely on the difference of the two rapidities associated to asymptotic states
\begin{equation}
\theta = \vartheta_1 - \vartheta_2 \ .
\end{equation}

The candidate expression for the one-loop S-matrix given in equation~\eqref{eqn:final} is then given by
\begin{align}
 {S^{(1)}}{}_{\IndA\IndB}^{\IndC\IndD}(\theta)=\frac{1}{8 \sinh\theta}\,\Big[{\tilde S^{(0)}}{}_{\IndA\IndB}^{\IndE\IndF}(\theta){\tilde S^{(0)}}{}_{\IndE\IndF}^{\IndC\IndD}(\theta)I(i\pi-\theta)\nonumber 
 &+ {\tilde S^{(0)}}{}_{\IndA\IndE}^{\IndF\IndC}(0){\tilde S^{(0)}}{}_{\IndF\IndB}^{\IndE\IndD}(\theta)I(0) \\ &
 +{\tilde S^{(0)}}{}_{\IndA\IndE}^{\IndF\IndD}(\theta){\tilde S^{(0)}}{}_{\IndF\IndB}^{\IndC\IndE}(\theta)I(\theta)\,\Big]
 \label{eqn:final_rel}~,
\end{align}
where $\tilde S{}^{(0)} (\theta)= 4 \sinh \theta \ S^{(0)}(\theta)$ and the one-loop integrals are given by
\begin{equation}\label{olrel}
I(\theta) = \frac{\theta\,\csch \theta}{4\pi i } \ .
\end{equation}
The consistency condition \eqref{consistency} now reads
\be\label{s2const}
{\tilde S^{(0)}}{}_{\IndA\IndE}^{\IndF\IndC}(0)\,{\tilde S^{(0)}}{}^{\IndE\IndD}_{\IndF\IndB}(\theta)\,=\,{\tilde S^{(0)}}{}^{\IndC\IndF}_{\IndA\IndE}(\theta)\,{\tilde S^{(0)}}{}^{\IndD\IndE}_{\IndF\IndB}(0)~.
\ee

\

As a starting point let us consider a general $\text{SO}(n)$-invariant theory with quartic interactions up to second order in derivatives for a single $\text{SO}(n)$ vector of unit mass $X$ (the light-cone derivatives $\partial_\pm$ are defined as $\partial_\tau \pm \partial_\sigma$)
\begin{equation}\label{s2laggen}
\mathcal{L} = \frac12 \partial_+ X \cdot \partial_- X -\frac12 X\cdot X + h \, \text{c}_1 (X\cdot X)^2 + h \, \text{c}_2 (X\cdot X)(\partial_+ X \cdot \partial_- X) + \ldots \ ,
\end{equation}
where we have eliminated the other allowed term $(X \cdot \partial_+ X)(X \cdot \partial_- X)$ using field redefinitions. $\text{c}_{1,2}$ are arbitrary constants, while $h$ is the small parameter that we use to do perturbation theory. We assume that the theory is renormalizable and the ellipses denote any higher-order interaction terms required therefor.

The requirements of $\text{SO}(n)$ invariance and crossing symmetry imply that the S-matrix can be parametrized in terms of two functions, $T(\theta)$ and $R(\theta)$, as follows
\begin{align}
& S_{ij}^{kl} (\theta) = \delta_{ij}\delta^{kl} \, R (i\pi - \theta) + \delta_i^k \delta_j^l \ T(\theta) + \delta_i^l \delta_j^k \ R(\theta) \ .\label{s2son}
\end{align}
where $i,j,\ldots=1,...,n$ are $\text{SO}(n)$ vector indices. Using standard perturbation theory with dimensional regularization~\footnote{To be clear, we use the $\overline{MS}$ scheme and drop the divergent pieces assuming they are either cancelled by the mass and wavefunction renormalization or absorbed into the renormalization of the coupling, which will not contribute here.} 
to one loop we find the following parametrizing functions
\begin{align}
T(\theta) &= 1 + 2 i h (\text{c}_1 + \text{c}_2) \csch \theta \no   \vphantom{\frac{2i h^2}{\pi}}
\\ & \quad \ \ \, + \frac{2ih^2}{\pi} \big(2 \text{c}_1 \text{c}_2 (i\pi - 2 \theta ) \coth \theta\csch\theta + i \pi  (2 \text{c}_1^2+2 \text{c}_1 \text{c}_2+\text{c}_2^2 (1+\cosh^2 \theta))  \csch^2\theta \no
\\ & \qquad \qquad \quad +((\text{c}_1 + \text{c}_2) (4 \text{c}_1+n (\text{c}_1+\text{c}_2))-2 \text{c}_2^2)  \csch\theta \big)\ ,  \vphantom{\frac{2i h^2}{\pi}}\no
\\
R(\theta) &= 2 i h (\text{c}_1 + \text{c}_2 \cosh\theta) \csch \theta  \vphantom{\frac{2i h^2}{\pi}}\label{s2com1}
\\ & \quad \ \ \, + \frac{2i h^2}{\pi}\big(\theta \, (\text{c}_1+\text{c}_2 \cosh \theta ) (\text{c}_1 (n+2)+\text{c}_2 (n-2) \cosh \theta ) \csch^2\theta  \no
\\ & \qquad \qquad \qquad + 2\pi i (\text{c}_1+\text{c}_2) (\text{c}_1+\text{c}_2 \cosh \theta ) \csch^2 \theta+  (2 \text{c}_1^2-\text{c}_2^2 \cosh \theta +\text{c}_2^2) \csch \theta\big)\ . \no \vphantom{\frac{2i h^2}{\pi}} 
\end{align}
Having checked that the tree-level S-matrix satisfies the consistency condition \eqref{s2const}, we can use the method of unitarity cuts described in Section \ref{sec:general} with $\IndA,\IndB,\ldots = i,j,\ldots$ to find the following result
\begin{align}
T_{\text{u.c.}}(\theta) &= 1 + 2 i h (\text{c}_1 + \text{c}_2) \csch \theta \no   \vphantom{\frac{2i h^2}{\pi}}
\\ & \quad \ \ \, + \frac{2ih^2}{\pi} \big(2 \text{c}_1 \text{c}_2 (i\pi - 2 \theta ) \coth \theta\csch\theta + i \pi  (2 \text{c}_1^2+2 \text{c}_1 \text{c}_2+\text{c}_2^2 (1+\cosh^2 \theta))  \csch^2\theta \no
\\ & \qquad \qquad \quad + (\text{c}_1+\text{c}_2) (4 \text{c}_1 + n (\text{c}_1+\text{c}_2))  \csch\theta \big) \ , \vphantom{\frac{2i h^2}{\pi}} \no
\\
R_{\text{u.c.}}(\theta) &= 2 i h (\text{c}_1 + \text{c}_2 \cosh\theta)  \csch \theta \vphantom{\frac{2i h^2}{\pi}} \label{s2com2}
\\ & \quad \ \ \, + \frac{2i h^2}{\pi}\big(\theta \, (\text{c}_1+\text{c}_2 \cosh \theta ) (\text{c}_1 (n+2)+\text{c}_2 (n-2) \cosh \theta ) \csch^2\theta \no
\\ & \qquad \qquad \qquad + 2\pi i (\text{c}_1+\text{c}_2) (\text{c}_1+\text{c}_2 \cosh \theta ) \csch^2 \theta+ 2 (\text{c}_1^2+\text{c}_2^2 \cosh \theta) \csch \theta\big) \ .\vphantom{\frac{2i h^2}{\pi}}\no
\end{align}
Clearly \eqref{s2com1} and \eqref{s2com2} differ. Furthermore, they only agree for $\text{c}_2 = 0$. Examining the Lagrangian \eqref{s2laggen} we see that this is precisely the situation in which there are no derivatives in the vertices, hence the one-loop bubble integrals in standard perturbation theory are finite. In fact, as one might expect, it can be seen that the difference between the two results comes precisely from rational terms that appear as a result of regularization.
Notice that the form of these terms is such that they are proportional to tree-level graphs. Therefore, introducing separate couplings $h_1=h c_1$ and $h_2=h c_2$ and shifting them in \eqref{s2laggen} as follows
\begin{align}\label{2shift}
 h_1 &\to h_1 + \frac{{h_2}^2}{\pi}& h_2&\to h_2 -\frac{3 {h_2}^2}{\pi}
\end{align}
we can recover the perturbative result. In this sense, the difference can be understood as a regularization-scheme ambiguity. On the other hand one may ask if performing unitarity cuts in $d = 2-2\epsilon$ dimensions \cite{vanNeerven:1985xr,Bern:1995db} could give rise to these rational terms and resolve this discrepancy. We leave the study of this formalism for future investigation.

In this work we would like to explore an alternative avenue, which is to focus on integrable theories. Integrable theories possess hidden symmetries that heavily constrain the scattering theory -- $(i)$ there can be no particle production, $(ii)$ the set of ingoing momenta should equal the set of outgoing momenta, and $(iii)$ the $n\to n$ scattering amplitude should factorize into a product of $2\to 2$ scattering amplitude \cite{Zamolodchikov:1978xm}. This final requirement implies the Yang-Baxter equation, a
consistency condition for equivalent orderings of scattering of three-particle states, which can be represented diagrammatically as follows:
\begin{equation}
\begin{tikzpicture}[line width = 2pt,scale=0.6,rotate=90,baseline=5.5em,yscale=-1]
\draw[->] (0,{16*(-5+4*sqrt(3))/23}) -- (8,{24*(12-5*sqrt(3))/23});
\node at (-0.4,{16*(-5+4*sqrt(3))/23-0.2}) {{\large $\vartheta_2$}};
\draw[->] ({16*(-5+4*sqrt(3))/23},0) -- ({24*(12-5*sqrt(3))/23},8);
\node at ({16*(-5+4*sqrt(3))/23},-0.45) {{\large $\vartheta_1$}};
\draw[->] ({8*(29-14*sqrt(3))/23},{24*(2+3*sqrt(3))/23}) -- ({24*(2+3*sqrt(3))/23},{8*(29-14*sqrt(3))/23});
\node at ({8*(29-14*sqrt(3))/23-0.3},{24*(2+3*sqrt(3))/23+0.3}) {{\large $\vartheta_3$}};
\node[circle,draw=black,fill=black!02,line width = 2pt,opacity=.90] at ({8*(7-sqrt(3))/23},{8*(7-sqrt(3))/23}) {{\Large $\theta_{12}$}};
\node[circle,draw=black,fill=black!10,line width = 2pt,opacity=.90] at ({8*(19-6*sqrt(3))/23},{8*(16+sqrt(3))/23}) {{\Large $\theta_{13}$}};
\node[circle,draw=black,fill=black!18,line width = 2pt,opacity=.90] at ({8*(16+sqrt(3))/23},{8*(19-6*sqrt(3))/23}) {{\Large $\theta_{23}$}};
\end{tikzpicture}
\qquad
\boldsymbol{=}
\qquad
\begin{tikzpicture}[line width = 2pt,scale=0.6,rotate=90,baseline=21.5em,yscale=-1]
\draw[->] (11,{8-24*(12-5*sqrt(3))/23}) -- (11+8,{8-16*(-5+4*sqrt(3))/23});
\node at (11-0.4,{8-24*(12-5*sqrt(3))/23-0.2}) {{\large $\vartheta_2$}}; 
\draw[->] ({11+8-24*(12-5*sqrt(3))/23},0) -- ({11+8-16*(-5+4*sqrt(3))/23},8);
\node at ({11+8-24*(12-5*sqrt(3))/23},-0.45) {{\large $\vartheta_1$}};
\draw[->] ({11+8-24*(2+3*sqrt(3))/23},{8-8*(29-14*sqrt(3))/23}) -- ({11+8-8*(29-14*sqrt(3))/23},{8-24*(2+3*sqrt(3))/23});
\node at ({11+8-24*(2+3*sqrt(3))/23-0.3},{8-8*(29-14*sqrt(3))/23+0.3}) {{\large $\vartheta_3$}}; 
\node[circle,draw=black,fill=black!02,line width = 2pt,opacity=.90] at ({11+8-8*(7-sqrt(3))/23},{8-8*(7-sqrt(3))/23}) {{\Large $\theta_{12}$}};
\node[circle,draw=black,fill=black!10,line width = 2pt,opacity=.90] at ({11+8-8*(19-6*sqrt(3))/23},{8-8*(16+sqrt(3))/23}) {{\Large $\theta_{13}$}};
\node[circle,draw=black,fill=black!18,line width = 2pt,opacity=.90] at ({11+8-8*(16+sqrt(3))/23},{8-8*(19-6*sqrt(3))/23}) {{\Large $\theta_{23}$}};
\end{tikzpicture}\label{ybe}
\end{equation}
As we will see in the following sections, for these models the unitarity techniques appear to work well giving the one-loop S-matrix associated to an ``integrable quantization'' up to finite shifts in the coupling. Furthermore, for theories that are also supersymmetric we find exact agreement.

\subsection{Integrable bosonic theories}\label{s31ibt}

Let us first discuss on a class of generalized sine-Gordon models \cite{Hollowood:1994vx,Bakas:1995bm}. These theories are defined by a gauged WZW model for a coset $G/H$ plus a potential and their classical integrability can be demonstrated through the existence of a Lax connection. Here we will consider the coset $G/H = \text{SO}(n+1)/\text{SO}(n)$, in which case the asymptotic excitations are a free $\text{SO}(n)$ vector with unit mass. Therefore, the S-matrix will again have the structure \eqref{s2son}. This class includes the sine-Gordon and complex sine-Gordon models for $n=1$ and $2$ respectively, 
for which the exact S-matrices are known \cite{Zamolodchikov:1978xm,Dorey:1994mg} and agree with perturbation theory.

Starting from the gauged WZW formulation of these models, the gauge symmetry can be fixed and at the classical level the unphysical modes integrated out \cite{Hoare:2010fb} giving a Lagrangian for just the physical excitations. To quartic order this is given by \eqref{s2laggen} with $\text{c}_1 = 0$ and $\text{c}_2 h = \tfrac{\pi}{2k}$, where $k$ is the coupling. The functions parametrizing the tree-level S-matrix are therefore
\begin{align}
T^{(0)}(\theta) = \frac{i\pi}{k}\csch \theta \ ,\qquad  R^{(0)}(\theta) = \frac{i\pi}{k}\coth \theta\ .\label{s2tlbos} 
\end{align}
For general $n$ this Lagrangian is only valid for the computation of the tree-level S-matrix -- the procedure of integrating out the unphysical fields picks up a one-loop correction \cite{Hoare:2010fb}
\begin{equation}
\Delta \mathcal{L} = -\frac{\pi}{2k^2} (X\cdot \partial_+ X)(X\cdot \partial_-X) - \frac{\pi(n-2)}{2k^2} (X\cdot X)(\partial_+ X \cdot \partial_- X) \ .\label{s2lagcon}
\end{equation}
Furthermore, the corresponding contributions to the one-loop S-matrix restore various properties of integrability. Indeed these counterterms were first studied from this perspective for the complex sine-Gordon model in \cite{deVega:1981ka}.

Including these contributions the resulting functions parametrizing the one-loop S-matrix following from perturbation theory with dimensional regularization are given
by~\footnote{For $n=1,2$ the S-matrix given by equations~\eqref{s2son},~\eqref{s2tlbos} and \eqref{s2olbos} satisfies the Yang-Baxter equation \eqref{ybe} to the appropriate order as expected. However, for $n\geq 3$ the situation is more subtle as the perturbative excitations appear as a limit of the kinks in the spectrum \cite{Hollowood:2010dt}. The S-matrix for the scattering of these kinks satisfies a dynamical Yang-Baxter equation \cite{Felder:1994be} whose semi-classical expansion gives a non-trivial modification of the usual classical Yang-Baxter equation \cite{Hoare:2013ysa}.}
\begin{align}
T^{(1)}(\theta) &= \frac{i\pi}{2k^2}  \big(i\pi  (1+\cosh^2\theta) \csch^2\theta - (n-2)\csch\theta\big)  \ ,\no
\\
R^{(1)}(\theta) &=
\frac{i\pi}{2k^2}  \big(2\pi i\coth\theta \csch \theta-2 (n-2) \coth \theta + \theta  (n-2) \coth^2 \theta \big)
 \ .  \label{s2olbos}
\end{align}

Inputting the tree-level S-matrix \eqref{s2tlbos} into the candidate formula for the one-loop S-matrix \eqref{eqn:final_rel} following from the unitarity techniques described in Section \ref{sec:general} we find the following expressions for the parametrizing functions
\begin{align}
T^{(1)}_{\text{u.c.}}(\theta) &= \frac{i\pi}{2k^2}  \big(i\pi  (1+\cosh^2\theta) \csch^2\theta + n\csch\theta\big)  \ ,\no
\\
R^{(1)}_{\text{u.c.}}(\theta) &= 
\frac{i\pi}{2k^2}  \big(2\pi i\coth\theta \csch \theta+2 \coth \theta + \theta  (n-2) \coth^2 \theta \big)\ . \label{s2olbosuc}
\end{align}
In this case these results agree with \eqref{s2olbos} up to a contribution proportional to tree-level S-matrix \eqref{s2tlbos}, which can be understood as just a shift of the coupling $k \to k + n - 1$. This is analogous to the shift mentioned in \eqref{2shift},  but it is not completely equivalent as here we are not comparing to standard perturbation theory, but rather to integrable S-matrix found from the gauged WZW model. It is important to observe that a consequence of this is that we do not need to introduce an additional coupling, and as such this can still be interpreted as a difference in the regularization scheme. 

For the sine-Gordon model ($n=1$) the S-matrix we are considering describes the scattering of a single particle type (the Lagrangian-field excitation) and therefore for the $2\to 2$ scattering is given by $T(\theta) + R(i\pi- \theta) + R(\theta)$. In this case the two results agree exactly. This should be expected as using field redefinitions the interaction piece of the sine-Gordon Lagrangian can be written in a form without derivatives.
For $n\geq 2$ the shift in the coupling is by the dual Coxeter number of the group $G=\text{SO}(n)$. This structure appears regularly in the quantization of WZW and gauged WZW models, where $k$ is the quantized level \cite{Witten:1983ar,Knizhnik:1984nr,Leutwyler:1991tv,Tseytlin:1992ri,Tseytlin:1993my,deWit:1993qv}. 

In summary, for a certain class of generalized sine-Gordon models we have found that the expression for the one-loop S-matrix gotten from unitarity techniques \eqref{eqn:final_rel} agrees with that found from standard perturbation theory (including the one-loop correction coming from the procedure of integrating out the unphysical fields \eqref{s2lagcon}) up to a scheme-dependent shift in the coupling.

\subsection{Integrable theories with fermions}

The models discussed in the previous section have an interesting origin in string theory. They appear as the Pohlmeyer reduction of strings moving on an $n+1$-sphere \cite{Pohlmeyer:1975nb,Eichenherr:1979yw}. In the string interpretation the reduction uses a non-local change of variables to solve the Virasoro constraints giving a classically equivalent theory \cite{Tseytlin:2003ii,Barbashov:1980kz,DeVega:1992xc}. The reduction can be extended to the Green-Schwarz action for the Type IIB superstring on AdS$_5 \times S^5$ \cite{Metsaev:1998it} giving a gauged WZW model for $\text{USp}(2,2)/\text{SU}(2)^2 \times \text{USp}(4)/\text{SU}(2)^2$ plus a potential and coupled to fermions \cite{Grigoriev:2007bu,Mikhailov:2007xr}. There are two truncations of this model that we will also study corresponding to the reduction of the superstring on AdS$_3 \times S^3$ \cite{Grigoriev:2008jq} and AdS$_2 \times S^2$ \cite{Grigoriev:2007bu}, 
by which we mean
the formal supercoset truncations of the full 10-d superstring theories on AdS$_3 \times S^3 \times T^4$ and on AdS$_2 \times S^2 \times T^6$ -- see, for example, \cite{Babichenko:2009dk} and \cite{Sorokin:2011rr} and references therein.

These reduced theories are all classically integrable, demonstrated by the existence of a Lax connection, and conjectured to be UV-finite \cite{Roiban:2009vh}. 
Indeed, in~\cite{Roiban:2009vh} finiteness at one loop and at two loops in the dimensional reduction scheme was demonstrated. Furthermore, the reduced AdS$_2 \times S^2$ theory is in fact given by the $\mathcal{N} = 2$ supersymmetric sine-Gordon model and hence is supersymmetric. The reduced AdS$_3 \times S^3$ and AdS$_5 \times S^5$ theories have a non-local $\mathcal{N}=4$ and $\mathcal{N}=8$ supersymmetry respectively \cite{Schmidtt:2010bi,Hollowood:2011fq,Goykhman:2011mq,Schmidtt:2011nr}, which manifests as a $q$-deformation of the S-matrix symmetry algebra.

The tree-level and one-loop S-matrices for these theories were computed in \cite{Hoare:2009fs,Hoare:2011fj}, while the exact S-matrices have been conjectured using integrability techniques in \cite{Kobayashi:1991rh} for the reduced AdS$_2 \times S^2$ model, \cite{Hoare:2011fj} for the reduced AdS$_3 \times S^3$ model and \cite{Hoare:2013ysa} for the reduced AdS$_5 \times S^5$ model.

In each of the reduced superstring theories the asymptotic excitations (both bosonic and fermionic) can be packaged into a single field 
\begin{equation} \label{superfield}
\Phi_{A\dot A} \,, \qquad A=(a|\alpha) \,, \qquad [a]=0 \,, \ [\alpha]=1 \ .
\end{equation}
The particular configurations relevant for the individual theories are then as follows:
\begin{equation}
\begin{array}{lll}
\textbf{Reduced AdS}\mathbf{_2 \times S^2:}  \qquad a=1\,, \ \alpha=2 &
\\\textbf{Reduced AdS}\mathbf{_3 \times S^3:}  \qquad a=1,2\,, \ \alpha=3,4 \quad & \text{and} \quad \Phi_{A\dot A\vphantom{\dot B}} = \Omega_{AB\vphantom{\dot B}}\Omega_{\dot A \dot B} \Phi_{B\dot B}
\\\textbf{Reduced AdS}\mathbf{_5 \times S^5:}  \qquad a=1,2\,, \ \alpha=3,4 & \end{array}
\end{equation}
where $\Omega_{ab} = \epsilon_{ab} \,, \ \Omega_{\alpha\beta} =\epsilon_{\alpha\beta}$ and $\Omega_{a\beta} = \Omega_{\alpha b} = 0$. In the reduced AdS$_5 \times S^5$ model the indices $a,\alpha,\dot a,\dot \alpha$ are $\text{SU}(2)$ fundamental indices, while in the reduced  AdS$_3 \times S^3$ model they are $\text{SO}(2)$ vector indices. 

In each case the global symmetry \cite{Hoare:2011fj} is such that the S-matrix should factorize under its structure~\footnote{This group-factorization property is exhibited by generic integrable theories with a non-simple global symmetry $G_1 \times G_2$, and the fields transforming in the bi-fundamental representation of this group~\cite{Ogievetsky:1987vv}\,\cite{Arutyunov:2009ga}. The global symmetries of the reduced AdS$_2 \times S^2$,  AdS$_3 \times S^3$ and AdS$_5 \times S^5$ models are discussed in \cite{Hoare:2011fj,Goykhman:2011mq,Hollowood:2011fq}.}  
\begin{equation}\label{s2smatfact}
S_{A\dot A,B\dot B}^{C \dot C, D \dot D}(\theta) = (-1)^{[\dot A][B]+[\dot C][D]} S_{AB\vphantom{\dot B}}^{CD \vphantom{\dot D}}(\theta) S_{\dot A \dot B}^{\dot C \dot D}(\theta)
\end{equation}
and indeed the tree-level results have this structure. Let us now present the tree-level S-matrices 
(we write only the undotted factor -- the dotted factor is given by the same expression) for each of the three reduced theories~\footnote{Compared to \cite{Hoare:2011fj} the following conventions have been changed: $(i)$ the r\^oles of the indices $a$ and $\alpha$ have been interchanged, which amounts to sending $k \to -k$ in the factor S-matrix, and $(ii)$ here we define $\epsilon_{12} = \epsilon^{12} = \epsilon_{34} = \epsilon^{34} = 1$.}
\allowdisplaybreaks{
\begin{align}
& \textbf{Reduced AdS}\mathbf{_2 \times S^2:} && 
S_{11}^{11} = M_1 \,, \
S_{22}^{22} = M_2 \,, \
S_{11}^{22} = M_3 \,, \
S_{22}^{11} = M_4 \,, \no
\\ & &&
S_{12}^{12} = M_5 \,, \
S_{21}^{21} = M_6 \,, \
S_{12}^{21} = M_7 \,, \
S_{21}^{12} = M_8 \,, \label{s222param}
\\\no
\\
& \textbf{Reduced AdS}\mathbf{_3 \times S^3:} && 
S_{AB}^{CD} = \left\{
\begin{array}{l} L_1 \delta_{ac}\delta_{bd}
              + L_2 \epsilon_{ac}\epsilon_{bd}
\\            L_3 \delta_{\alpha\gamma}\delta_{\beta\delta}
              + L_4 \epsilon_{\alpha\gamma}\epsilon_{\beta\delta}
\\             L_5 \delta_{ac}\delta_{\beta\delta}
              + L_6 \epsilon_{ac}\epsilon_{\beta\delta}
\\             L_7 \delta_{\alpha\gamma}\delta_{bd}
              + L_8 \epsilon_{\alpha\gamma}\epsilon_{bd}
\\             L_{9} (\delta_{ab} \delta_{\gamma\delta}
                     +\epsilon_{ab} \epsilon_{\gamma\delta})
\\              L_{10} (\delta_{\alpha\beta} \delta_{cd}
                     +\epsilon_{\alpha\beta} \epsilon_{cd})
\\             L_{11} (\delta_{ad} \delta_{\gamma\beta}
                      +\epsilon_{ad} \epsilon_{\gamma\beta})
\\                L_{12} (\delta_{\alpha\delta} \delta_{cb}
                       +\epsilon_{\alpha\delta} \epsilon_{cb})
\end{array}\right. \label{s233param}
\\\no
\\
& \textbf{Reduced AdS}\mathbf{_5 \times S^5:} && 
S_{AB}^{CD} =
\left\{
\begin{array}{l} K_1 \delta_a^c\delta_b^d
              + K_2 \delta_a^d\delta_b^c
\\             K_3 \delta_\alpha^\gamma\delta_\beta^\delta  
              + K_4 \delta_\alpha^\delta\delta_\beta^\gamma
\\             K_5 \epsilon_{ab}\epsilon^{\gamma\delta}
            \hspace{9pt} K_6 \epsilon_{\alpha\beta}\epsilon^{cd}
\\             K_7  \delta_a^d \delta_\beta^\gamma
           \hspace{15pt} K_8 \delta_\alpha^\delta \delta_b^c
\\             K_9 \delta_a^c \delta_\beta^\delta
            \hspace{16pt} K_{10} \delta_\alpha^\gamma \delta_b^d
\end{array} \right. \label{s255param}
\end{align}
}
where
\begin{align}
\no
&  M^{(0)}_1 (\theta) = - M^{(0)}_2 (\theta) = \frac{i\pi}k \csch\theta
&& M^{(0)}_3 (\theta) = M^{(0)}_4 (\theta) = -\frac{i\pi}{2k} \sech \frac\theta 2
\\
&  M^{(0)}_5 (\theta) = - M^{(0)}_6 (\theta) = 0
&& M^{(0)}_7 (\theta) = M^{(0)}_8 (\theta) = \frac{i\pi}{2k}\csch \frac \theta 2
\label{s222func}
\\\no
\\
\no
&  L^{(0)}_1 (\theta) = - L^{(0)}_3 (\theta) = \frac{i\pi}k \csch \theta 
&& L^{(0)}_2 (\theta) = - L^{(0)}_4 (\theta) = - \frac{i\pi}k \coth \theta
\\
\no
&  L^{(0)}_5 (\theta) = - L^{(0)}_7 (\theta) = 0
&&  L^{(0)}_6 (\theta) = - L^{(0)}_8 (\theta) = 0
\\
&  L^{(0)}_9 (\theta) = L^{(0)}_{10} (\theta) = -\frac{i\pi}{2k} \sech \frac \theta 2
&&  L^{(0)}_{11} (\theta) = L^{(0)}_{12} (\theta) = \frac{i\pi}{2k}\csch \frac \theta 2
\label{s233func}
\\\no
\\
\no
&  K^{(0)}_1 (\theta) = K^{(0)}_3 (\theta) = -\frac{i\pi}{2k}\tanh\frac\theta 2
&& K^{(0)}_2 (\theta) = K^{(0)}_4 (\theta) = \frac{i\pi}k\coth\theta
\\
\no
&  K^{(0)}_5 (\theta) = K^{(0)}_6 (\theta) = -\frac{i\pi}{2k}\sech\frac\theta 2
&& K^{(0)}_7 (\theta) = K^{(0)}_8 (\theta) = \frac{i\pi}{2k}\csch\frac\theta 2
\\
&  K^{(0)}_9 (\theta) = K^{(0)}_{10} (\theta) = 0
&&
\label{s255func}
\end{align}

Due to the presence of fermionic fields the consistency condition \eqref{s2const} is generalized to
\be\label{s2const_ferm}
(-1)^{[\IndC][\IndF]+[\IndE][\IndF]}{\tilde S^{(0)}}{}_{\IndA\IndE}^{\IndF\IndC}(0)\,{\tilde S^{(0)}}{}^{\IndE\IndD}_{\IndF\IndB}(\theta)\,=\,(-1)^{[\IndD][\IndE]+[\IndE][\IndF]}{\tilde S^{(0)}}{}^{\IndC\IndF}_{\IndA\IndE}(\theta)\,{\tilde S^{(0)}}{}^{\IndD\IndE}_{\IndF\IndB}(0)~.
\ee
and indeed the tree-level S-matrices \eqref{s2smatfact}--\eqref{s255func}, both the full, $S_{A\dot A,B\dot B}^{C\dot C D \dot D}$ ($M,N,\ldots = (A,\dot A),\,(B,\dot B)\ldots$), and factor, $S_{AB}^{CD}$ ($M,N,\ldots = A,B,\ldots$), satisfy this relation. We can therefore input them into the candidate expression for the one-loop S-matrix derived from the procedure described in Section \ref{sec:general}
\begin{equation}\begin{split}
 {S^{(1)}}{}_{\IndA\IndB}^{\IndC\IndD}(\theta)=\frac{1}{8 \sinh \theta}\,\Big[&{\tilde S^{(0)}}{}_{\IndA\IndB}^{\IndE\IndF}(\theta){\tilde S^{(0)}}{}_{\IndE\IndF}^{\IndC\IndD}(\theta)I(i\pi-\theta)\\
  \vphantom{\frac{1}{4 (\e_2\,\rmp_1-\e_1\,\rmp_2)}}
 +&  (-1)^{[\IndC][\IndF]+[\IndE][\IndF]}\,{\tilde S^{(0)}}{}_{\IndA\IndE}^{\IndF\IndC}(0){\tilde S^{(0)}}{}_{\IndF\IndB}^{\IndE\IndD}(\theta)I(0) \\ 
 \vphantom{\frac{1}{4 (\e_2\,\rmp_1-\e_1\,\rmp_2)}}
 +&(-1)^{[\IndC][\IndE]+[\IndD][\IndF]+[\IndE][\IndF]+[\IndC][\IndD]}{\tilde S^{(0)}}{}_{\IndA\IndE}^{\IndF\IndD}(\theta){\tilde S^{(0)}}{}_{\IndF\IndB}^{\IndC\IndE}(\theta)I(\theta)\,\Big]~,\label{eqn:final_ferm_rel}
\end{split}\end{equation}
where $\tilde S{}^{(0)} (\theta)= 4 \sinh \theta \ S^{(0)}(\theta)$ and the one-loop integrals are given in \eqref{olrel}. Again this can be done with both the full ($M,N,\ldots = (A,\dot A),\,(B,\dot B)\ldots$) and factor ($M,N,\ldots = A,B,\ldots$) tree-level S-matrices. As expected they give the same one-loop S-matrix parametrized by the following functions
\begin{align}
\no
&  M^{(1)}_1 (\theta) = M^{(1)}_2 (\theta) = \frac{P(\theta)}2
&& M^{(1)}_3 (\theta) = M^{(1)}_4 (\theta) = 0
\\
&  M^{(1)}_5 (\theta) = M^{(1)}_6 (\theta) = \frac{P(\theta)}2
&& M^{(1)}_7 (\theta) = M^{(1)}_8 (\theta) = 0
\label{s222funcol}
\\\no
\\
\no
&  L^{(1)}_1 (\theta) = L^{(1)}_3 (\theta) = - \frac{\pi^2}{2k^2} + P(\theta)
&& L^{(1)}_2 (\theta) = L^{(1)}_4 (\theta) = \tilde P(\theta)
\\
\no
&  L^{(1)}_5 (\theta) = L^{(1)}_7 (\theta) = P(\theta)
&&  L^{(1)}_6 (\theta) = L^{(1)}_8 (\theta) = \tilde P(\theta)
\\
&  L^{(1)}_9 (\theta) = L^{(1)}_{10} (\theta) = 0
&&  L^{(1)}_{11} (\theta) = L^{(1)}_{12} (\theta) = 0
\label{s233funcol}
\\\no
\\\no
&  K^{(1)}_1 (\theta) = - K^{(1)}_3 (\theta) = -\frac{5\pi^2}{8k^2} -\frac{i\pi \theta}{2k^2}  + \frac{P(\theta)}2
&& K^{(1)}_2 (\theta) = - K^{(1)}_4 (\theta) = \frac{\pi^2}{2k^2} + \frac{i\pi\theta}{k^2}
\\
\no
&  K^{(1)}_5 (\theta) = K^{(1)}_6 (\theta) = 0
&& K^{(1)}_7 (\theta) = K^{(1)}_8 (\theta) = 0
\\
&  K^{(1)}_9 (\theta) = - K^{(1)}_{10} (\theta) = \frac{P(\theta)}2
&&
\label{s255funcol}
\end{align}
\begin{align}\no
& P(\theta) = \frac{i\pi}{k^2}\csch\theta + \frac{i\pi}{2k^2}(i\pi-2\theta)\coth\theta\csch\theta - \frac{\pi^2}{2k^2}\csch^2\theta 
\\\no
& \tilde P(\theta) = - \frac{i\pi}{k^2} \coth \theta - \frac{i\pi}{2k^2}(i\pi-2\theta)\csch^2\theta + \frac{\pi^2}{2k^2}\coth\theta\csch\theta\ .
\end{align}
These functions are in exact agreement with those found by perturbation theory \cite{Hoare:2011fj}. In contrast to the bosonic theories discussed in Section \ref{s31ibt} no additional shift of the coupling is required. The presence of the supersymmetry, albeit deformed, may provide an explanation for this, with shifts arising from bosonic loops cancelled by shifts from fermionic loops. 
Indeed, this is a feature of supersymmetric WZW and gauged WZW theories -- see, for example, \cite{Tseytlin:1993my}. Furthermore, we have also checked that the unitarity-cutting procedure matches the perturbative result at one-loop in the $\mathcal{N}=1$ supersymmetric sine-Gordon model \cite{Shankar:1977cm}.

Let us mention that in the reduced AdS$_3 \times S^3$ standard perturbative computation a contribution coming from a one-loop correction needs to be added so that the S-matrix satisfies the Yang-Baxter equation. It is this S-matrix that the unitarity technique matches. 
This is in direct analogy with the story for the bosonic models discussed in Section \ref{s31ibt}, except that currently no path integral origin for the correction in the AdS$_3 \times S^3$ case is known. Together with the complex sine-Gordon case this is another example of how unitarity methods applied to a classically integrable theory seem to provide a quantum integrable result. This seems to suggest a relationship between integrable quantization and unitarity techniques which would be interesting to investigate further. 

\section{String theory}
\label{sec:strings}

With the strong indication that in the presence of integrability and supersymmetry the method of unitarity cuts gives the correct result for the one-loop S-matrix we move onto the case of the light-cone gauge-fixed superstring on AdS$_5 \times S^5$ and the world-sheet S-matrix.

The integrability of the world-sheet sigma model is a well-established statement~\cite{Bena:2003wd}\,\cite{Kazakov:2004qf,Berkovits:2004xu} at the classical level.  Assuming the quantum integrability of the full world-sheet theory and using the global symmetries the \emph{exact} world-sheet S-matrix has been uniquely determined~\cite{Beisert:2005tm} up to an overall phase, or dressing factor~\cite{Arutyunov:2004vx}. The determination of the latter exploited the non-relativistic generalization of the crossing symmetry~\cite{Janik:2006dc,Volin:2009uv}
as well as perturbative data both from the string and gauge theory sides~\cite{Beisert:2006ib,Beisert:2006ez}. 
For a comprehensive reviews and further references see~\cite{Arutyunov:2009ga,Beisert:2010jr}. 

The perturbative study of the two-body S-matrix for the world-sheet sigma-model  (for a review, see \cite{Arutyunov:2009ga,McLoughlin:2010jw}) was initiated in~\cite{Klose:2006zd}~\footnote{Earlier work on related models  with truncated field content appeared in~\cite{Klose:2006dd,Roiban:2006yc}.} starting from the Green-Schwarz action in the so-called generalized uniform light-cone gauge~\cite{Arutyunov:2006ak,Arutyunov:2006gs,Frolov:2006cc} and applying LSZ reduction to its quartic vertices. Due to gauge-fixing the theory does \emph{not} possess world-sheet Lorentz invariance, however the off-shell symmetry algebra is $\mathfrak{psu}(2|2)^2 \ltimes \mathbb{R}^3$, which originates from the $\mathfrak{psu}(2,2|4)$ target-space symmetry of the Green-Schwarz action \cite{Metsaev:1998it}. The action of this symmetry is non-local and the central extensions encode the 2-d energy and momentum of the theory~\cite{Arutyunov:2006ak}. It can therefore, in some sense, be understood 
as a non-relativistic non-local generalization of world-sheet supersymmetry. 
Furthermore, while the theory is not power-counting renormalizable, 
it is believed to be UV-finite -- for an extensive discussion of related issues see, for example, \cite{Roiban:2007jf}. 

Relaxing the level-matching condition and taking the limit of infinite light-cone momentum (decompactification limit), the world-sheet theory becomes a massive field theory defined on a plane, with well-defined asymptotic states and S-matrix. 
The scattering of the world-sheet excitations has been studied at tree-level in~\cite{Klose:2006zd}, while one-loop~\cite{Klose:2007wq} and two-loop~\cite{Klose:2007rz} results have been carried out only in the simpler near-flat-space limit~\cite{Maldacena:2006rv} where interactions are at most quartic in the fields. These studies have also explicitly shown some consequences of the integrability of the model, such as the factorization of the many-body S-matrix  and the absence of particle production in the scattering processes~\cite{Puletti:2007hq}.

With relaxed level-matching, the symmetry group is a centrally extended PSU$(2|2)\times$ PSU$(2|2)$~\cite{Arutyunov:2006ak}, the same appearing in the dual gauge theory~\cite{Beisert:2005tm}. The charges of the fields under the bosonic subalgebra SU(2)$^4$ can again be indicated via the double-index notation \eqref{superfield}, with $a,\alpha,\dot a,\dot \alpha$ being $\text{SU}(2)$ fundamental indices combined as $A=(a|\alpha),\dot A=(\dot a|\dot\alpha)$  into the fundamental indices of the two PSU$(2|2)$ factors. The global symmetry structure should lead to a non-relativistic generalization of the group-factorization~\footnote{This can also be interpreted as  the requirement that the Faddeev-Zamolodchikov algebra, used in describing the Hilbert space of the asymptotic states, is a direct product~\cite{Arutyunov:2006yd,Klose:2006zd}.}  \eqref{s2smatfact} 
\begin{equation}\label{s2smatfact_string}
S_{A\dot A,B\dot B}^{C \dot C, D \dot D}(p_1,p_2) = (-1)^{[\dot A][B]+[\dot C][D]} S_{AB\vphantom{\dot B}}^{CD \vphantom{\dot D}}(p_1,p_2) S_{\dot A \dot B}^{\dot C \dot D}(p_1,p_2)~,
\end{equation}
which has indeed been verified at the tree level~\cite{Klose:2006zd}.

Since only the SU$(2)^2$ of each PSU$(2|2)$  is manifest in the gauge-fixed world-sheet theory, the tree-level S-matrices  are parametrized as follows  in terms of the basic SU$(2)$-invariants 
\begin{equation}\label{Sinvariants}
S_{AB}^{CD} = \left\{\begin{array}{l} A \delta_a^c \delta_b^d + B \delta_a^d \delta_b^c
\\                                     D \delta_\alpha^\gamma \delta_\beta^\delta + E \delta_\alpha^\delta \delta_\beta^\gamma
\\                                     C \epsilon_{ab}\epsilon^{\gamma \delta} \quad \ \, F \epsilon_{\alpha\beta}\epsilon^{cd}
\\                                     G \delta_a^c \delta_\beta^\delta \qquad H \delta_a^d \delta_\beta^\gamma
\\                                     L \delta_\alpha^\gamma \delta_b^d \qquad K \delta_\alpha^\delta \delta_b^c \end{array}\right. ~.
\end{equation}
The functions above were obtained in the generalized uniform light-cone gauge and therefore they show an explicit dependence on the parameter $a$ labeling different light-cone gauge choices~\cite{Arutyunov:2006gs}. In~\cite{Klose:2006zd} those functions were evaluated at leading order in perturbation theory, where the small parameter $\co$ is the inverse of the string tension 
\begin{equation}
\co^{-1} = g = \frac{\sqrt{\lambda}}{2\pi} \ .
\end{equation} 
Here we present the free part (given by the identity operator) and the tree-level expressions for the functions above
\ba
\!\!A^{(\rm free)} &=& D^{(\rm free)}~~ =~~ G^{(\rm free)}~~ =~~ L^{(\rm free)}~~ =~~ 1\nonumber\\ 
\!\!  B^{(\rm free)}& =& C^{(\rm free)} ~~=~~ E^{(\rm free)}~~ = ~~F^{(\rm free)}~~ = ~~H^{(\rm free)} ~~=~~ K^{(\rm free)}~~ =~~ 0\ ,\\
\!\!A^{(0)} & =& -\frac{i\co}2 (\epsilon_2 \rmp_1 - \epsilon_1 \rmp_2) (a-\tfrac12) A^{(\rm free)} + \frac{i\co}4 \frac{(\rmp_1-\rmp_2)^2}{\epsilon_2 \rmp_1 - \epsilon_1 \rmp_2} \ , \nonumber
\\ \nonumber
\!\!B^{(0)} & =& -\frac{i\co}2 (\epsilon_2 \rmp_1 - \epsilon_1 \rmp_2) (a-\tfrac12) B^{(\rm free)} + i\co \frac{\rmp_1 \rmp_2}{\epsilon_2 \rmp_1 - \epsilon_1 \rmp_2} \ ,
\\ \nonumber
\!\!D^{(0)} & = &-\frac{i\co}2 (\epsilon_2 \rmp_1 - \epsilon_1 \rmp_2) (a-\tfrac12) D^{(\rm free)} - \frac{i\co}4 \frac{(\rmp_1-\rmp_2)^2}{\epsilon_2 \rmp_1 - \epsilon_1 \rmp_2} \ , 
\\
\!\!E^{(0)} & = &-\frac{i\co}2 (\epsilon_2 \rmp_1 - \epsilon_1 \rmp_2) (a-\tfrac12) E^{(\rm free)} - i\co \frac{\rmp_1 \rmp_2}{\epsilon_2 \rmp_1 - \epsilon_1 \rmp_2} \ , \nonumber
\\
\!\!C^{(0)} & =& -\frac{i\co}2 (\epsilon_2 \rmp_1 - \epsilon_1 \rmp_2) (a-\tfrac12) C^{(\rm free)} + \frac{i\co}2 \sqrt{(\epsilon_1+1)(\epsilon_2+1)} \frac{\epsilon_2 \rmp_1 - \rmp_2 \epsilon_1 - \rmp_1 + \rmp_2}{\epsilon_2 \rmp_1 - \epsilon_1 \rmp_2} \ , \nonumber
\\
\!\!F^{(0)} & = &-\frac{i\co}2 (\epsilon_2 \rmp_1 - \epsilon_1 \rmp_2) (a-\tfrac12) F^{(\rm free)} + \frac{i\co}2 \sqrt{(\epsilon_1+1)(\epsilon_2+1)} \frac{\epsilon_2 \rmp_1 - \rmp_2 \epsilon_1 - \rmp_1 + \rmp_2}{\epsilon_2 \rmp_1 - \epsilon_1 \rmp_2} \ , \nonumber
\\
\!\!H^{(0)} & =& -\frac{i\co}2 (\epsilon_2 \rmp_1 - \epsilon_1 \rmp_2) (a-\tfrac12) H^{(\rm free)} + \frac{i\co}2 \frac{\rmp_1 \rmp_2}{\epsilon_2 \rmp_1 - \epsilon_1 \rmp_2} \frac{(\epsilon_1+1)(\epsilon_2+1)- \rmp_1\rmp_2}{\sqrt{(\epsilon_1+1)(\epsilon_2+1)}} \ , \nonumber
\\
\!\!K^{(0)} & =& -\frac{i\co}2 (\epsilon_2 \rmp_1 - \epsilon_1 \rmp_2) (a-\tfrac12) K^{(\rm free)} + \frac{i\co}2 \frac{\rmp_1 \rmp_2}{\epsilon_2 \rmp_1 - \epsilon_1 \rmp_2} \frac{(\epsilon_1+1)(\epsilon_2+1)- \rmp_1\rmp_2}{\sqrt{(\epsilon_1+1)(\epsilon_2+1)}} \ , \nonumber
\\
\!\!G^{(0)} & =& -\frac{i\co}2 (\epsilon_2 \rmp_1 - \epsilon_1 \rmp_2) (a-\tfrac12) G^{(\rm free)} - \frac{i\co}4 \frac{\rmp_1^2 - \rmp_2^2}{\epsilon_2 \rmp_1 - \epsilon_1 \rmp_2} \ , \nonumber
\\
\!\!L^{(0)} & =& -\frac{i\co}2 (\epsilon_2 \rmp_1 - \epsilon_1 \rmp_2) (a-\tfrac12) L^{(\rm free)} + \frac{i\co}4 \frac{\rmp_1^2 - \rmp_2^2}{\epsilon_2 \rmp_1 - \epsilon_1 \rmp_2} \ .
\ea
Above, $\e_i=\sqrt{1+\rmp_i^2}$ is the relativistic energy, which follows from the non-relativistic dispersion relation $\epsilon(\rmp)=\sqrt{1+\frac{\lambda}{\pi^2}\sin^2\frac{\rmp}{2}}$~\cite{Beisert:2004hm,Beisert:2005tm} expanded in the near-BMN limit, $\rmp \to \co \rmp$, corresponding to the perturbative regime.
After having explicitly verified that the matrix elements above verify the consistency relation \eqref{consistency_ferm}, we can safely use them in the expression \eqref{eqn:final_ferm} with \eqref{bubblep}-\eqref{bubblem} and get the one-loop S-matrix for the  light-cone gauge-fixed sigma model.
The result can be written as follows 
\begin{equation}\begin{split}\label{Sstrings_cut}
S_{AB}^{CD}(\rmp_1,\rmp_2) & = \exp\big(i\varphi_a(\rmp_1,\rmp_2)\big) \ \tilde S_{AB}^{CD} \\ & = \exp\big(-\tfrac{i\co}2(e_2\rmp_1 - e_1\rmp_2)(a-\tfrac12) + i\co^2 \tilde\varphi (\rmp_1,\rmp_2)\big) \ \tilde S_{AB}^{CD} + \mathcal{O}(\co^3) \ , 
\end{split}\end{equation}
where we have pulled out a factor that to the one-loop order can be resummed as an overall phase (this exponentiation is consistent with the requirement of integrability that all dynamical information and the gauge dependence on the parameter $a$ should be encoded in the scalar factor~\cite{Arutyunov:2006iu}). 
The remaining part $\tilde S_{AB}^{CD}$ has the same structure as in \eqref{Sinvariants} 
with parametrizing functions to the one-loop order given by
\allowdisplaybreaks{
\begin{align}\nonumber
\tilde A^{(1)} & = 1 + \frac{i\co}4 \frac{(\rmp_1-\rmp_2)^2}{\epsilon_2 \rmp_1 - \epsilon_1 \rmp_2} + \frac{\co^2}{4} \big(\rmp_1\rmp_2-\frac{(\rmp_1+\rmp_2)^4}{8(\epsilon_2 \rmp_1 - \epsilon_1 \rmp_2)^2}\big) \ ,
\\\nonumber
\tilde B^{(1)} & = i\co \frac{\rmp_1 \rmp_2}{\epsilon_2 \rmp_1 - \epsilon_1 \rmp_2} - \frac{\co^2}4 \rmp_1 \rmp_2 \ ,
\\\nonumber
\tilde D^{(1)} & = 1 - \frac{i\co}4 \frac{(\rmp_1-\rmp_2)^2}{\epsilon_2 \rmp_1 - \epsilon_1 \rmp_2} + \frac{\co^2}{4} \big(\rmp_1\rmp_2-\frac{(\rmp_1+\rmp_2)^4}{8(\epsilon_2 \rmp_1 - \epsilon_1 \rmp_2)^2}\big) \ ,
\\\nonumber
\tilde E^{(1)} & = - i\co \frac{\rmp_1 \rmp_2}{\epsilon_2 \rmp_1 - \epsilon_1 \rmp_2} - \frac{\co^2}4 \rmp_1 \rmp_2 \ ,
\\\nonumber
\tilde C^{(1)} & = \frac{i\co}2 \sqrt{(\epsilon_1+1)(\epsilon_2+1)} \frac{\epsilon_2 \rmp_1 - \rmp_2 \epsilon_1 - \rmp_1 + \rmp_2}{\epsilon_2 \rmp_1 - \epsilon_1 \rmp_2} \ ,
\\\nonumber
\tilde F^{(1)} & = \frac{i\co}2 \sqrt{(\epsilon_1+1)(\epsilon_2+1)} \frac{\epsilon_2 \rmp_1 - \rmp_2 \epsilon_1 - \rmp_1 + \rmp_2}{\epsilon_2 \rmp_1 - \epsilon_1 \rmp_2} \ ,
\\\nonumber
\tilde H^{(1)} & = \frac{i\co}2 \frac{\rmp_1 \rmp_2}{\epsilon_2 \rmp_1 - \epsilon_1 \rmp_2} \frac{(\epsilon_1+1)(\epsilon_2+1)- \rmp_1\rmp_2}{\sqrt{(\epsilon_1+1)(\epsilon_2+1)}} \ ,
\\\nonumber
\tilde K^{(1)} & = \frac{i\co}2 \frac{\rmp_1 \rmp_2}{\epsilon_2 \rmp_1 - \epsilon_1 \rmp_2} \frac{(\epsilon_1+1)(\epsilon_2+1)- \rmp_1\rmp_2}{\sqrt{(\epsilon_1+1)(\epsilon_2+1)}} \ ,
\\\nonumber
\tilde G^{(1)} & = 1 - \frac{i\co}4 \frac{\rmp_1^2 - \rmp_2^2}{\epsilon_2 \rmp_1 - \epsilon_1 \rmp_2} + \frac{\co^2}{8} \big(\rmp_1\rmp_2-\frac{(\rmp_1+\rmp_2)^4}{4(\epsilon_2 \rmp_1 - \epsilon_1 \rmp_2)^2}\big) \ ,
\\\label{functions_strings_cuts}
\tilde  L^{(1)} & = 1 + \frac{i\co}4 \frac{\rmp_1^2 - \rmp_2^2}{\epsilon_2 \rmp_1 - \epsilon_1 \rmp_2} + \frac{\co^2}{8} \big(\rmp_1\rmp_2-\frac{(\rmp_1+\rmp_2)^4}{4(\epsilon_2 \rmp_1 - \epsilon_1 \rmp_2)^2}\big) \ ,
\end{align}}
and 
\begin{equation}\label{phase_strings_cuts}
\tilde \varphi(\rmp_1,\rmp_2) = \frac{1}{2\pi}\frac{\rmp_1^2 \rmp_2^2 \big((\epsilon_2 \rmp_1 - \epsilon_1 \rmp_2) - (\epsilon_1 \epsilon_2 - \rmp_1 \rmp_2)\arsinh [\epsilon_2 \rmp_1 - \epsilon_1 \rmp_2] \big)}{(\epsilon_2 \rmp_1 - \epsilon_1 \rmp_2)^2}\ . 
\end{equation}

As mentioned above, because of the complicated structure of interactions of the light-cone gauge-fixed sigma model,
the perturbative S-matrix is known beyond the leading order~\cite{Klose:2007wq,Klose:2007rz} only in the   kinematic truncation 
known as near-flat-space limit~\cite{Maldacena:2006rv}.
Therefore, to test the validity of the unitarity method,  we need to compare our one-loop result to the corresponding limit of the exact world-sheet S-matrix. 
This is achieved by extending the analysis of~\cite{Klose:2006zd} to next-to-leading order, where the comparison between the perturbative S-matrix and the exact one was performed at the tree level. One considers the matrix elements derived in~\cite{Beisert:2005tm} for a single SU$(2|2)$ sector together with the dressing phase, here needed at next-to-leading order in the $1/\sqrt{\lambda}$ expansion. In the comparison with the world-sheet calculation all dimensional quantities (such as the spin-chain length and the momenta) should be rescaled via a factor of $\sqrt{\lambda}/(2\pi)$~\cite{Klose:2006zd}, for us $\rmp\to \co\,\rmp$. 

\def\bes{{\bf B}}
Here we take the form of the matrix elements of~\cite{Beisert:2005tm} given in eq.~(6.9) of \cite{Klose:2006zd}. To be explicit let us define
\begin{equation}\begin{split}
A_{\rm ex} = & \frac1{2\sqrt{A^\bes}}(A^\bes - B^\bes) \ , \qquad B_{\rm ex} = \frac1{2\sqrt{A^\bes}}(A^\bes + B^\bes) \ , \qquad C_{\rm ex} = \frac1{2\sqrt{A^\bes}}C^\bes \ ,
\\
D_{\rm ex} = & \frac1{2\sqrt{A^\bes}}(-D^\bes + E^\bes) \ , \quad \, E_{\rm ex} = \frac1{2\sqrt{A^\bes}}(-D^\bes - E^\bes) \ , \quad  \,F_{\rm ex} = - \frac1{2\sqrt{A^\bes}}F^\bes \ ,
\\
H_{\rm ex} = & \frac1{\sqrt{A^\bes}}H^\bes \ , \qquad
K_{\rm ex} = \frac1{\sqrt{A^\bes}}K^\bes \ , \qquad
G_{\rm ex} = \frac1{\sqrt{A^\bes}}G^\bes \ , \qquad
L_{\rm ex} = \frac1{\sqrt{A^\bes}}L^\bes \ ,
\end{split}\end{equation}
where $A_{\rm ex},\ldots,L_{\rm ex}$ are comparable to the parametrizing functions $A,\ldots,L$ \eqref{Sinvariants}.
The final piece of information required is the phase $e^{i\theta(\rmp_1,\rmp_2)}$ (defined as in eq.~(6.12) of \cite{Klose:2006zd}). The leading order piece of $\theta(\rmp_1,\rmp_2)$ is given in eq.~(6.13) of \cite{Klose:2006zd}, while at next-to-leading order we found it useful to use the expression given in eqs.~(15)-(19) of \cite{Arutyunov:2006iu}. Expanding the phase in the near-BMN limit gives
\begin{equation}
\exp\big(i\theta(\rmp_1,\rmp_2)\big) = \exp \Big(i\co\, \frac{(\rmp_1-\rmp_2-\e_2\rmp_1+\e_1\rmp_2)^2}{2(\epsilon_2\rmp_1 - \epsilon_1\rmp_2)}
 + 2i\co^2\tilde \varphi(\rmp_1,\rmp_2)+\mathcal{O}(\co^3) \Big)\ ,
\end{equation}
where $\tilde \varphi(\rmp_1,\rmp_2)$ is defined in eq.~\eqref{phase_strings_cuts}.

The exact S-matrix then should be compared with the string calculation in the constant-$J$ gauge $a = 0$. Doing so we find
\begin{equation}\label{strings_compar_2}
({{S}}_{AB}^{CD})_{\rm ex}= 
e^{\frac{i\co}4 \big(([A] + 2[B] -[C] - 2) \rmp_1 +([B] - 2[C] - [D]+2) \rmp_2\big)} \,e^{\varphi_{a=0}(\rmp_1,\rmp_2)} \, \tilde S_{AB}^{CD} + \mathcal{O}(\co^3) \ . 
\end{equation}
From \eqref{strings_compar_2} we see that we have agreement up to a phase whose argument is linear in momenta.
This is not surprising, as it simply amounts to moving from the string frame to the spin-chain frame~\cite{Arutyunov:2006yd,Ahn:2010ka}. As argued already at the tree level~\cite{Klose:2006zd} such terms should not
affect the physical spectrum following from inputting the S-matrix into the asymptotic Bethe equations. 

In summary, up to a phase whose argument is linear in momenta, the method of unitarity cuts reproduces the near-BMN expansion of the  string world-sheet S-matrix~\cite{Beisert:2005tm}. It is important to note that the result we are comparing to is that found using the techniques of integrability.
Indeed, previous attempts have been made to compute the one-loop result using standard perturbation
theory, however there are unresolved issues relating to regularization (see footnote \ref{foot1}). As discussed in Section \ref{sec:general}, assuming there exists a suitable regularization scheme and there are no additional
rational pieces (as is apparently the case for the string world-sheet S-matrix), our procedure na\"ively circumvents these problems
as \eqref{eqn:final} and \eqref{eqn:final_ferm} are manifestly finite.

\section{Conclusions}
\label{sec:conclusions}

In this work we have applied the method of unitarity cuts to two-dimensional quantum field theories. 
The computation of the cut-constructible piece of the one-loop four-point scattering amplitude (from which follows the S-matrix describing the scattering of two particles) reduces to a sum of products of two tree-level amplitudes weighted by scalar bubble integrals. 

As in four dimensions, it is not immediately clear in which theories the cut-constructible piece provides the full result -- that is there are no rational terms. The examples studied in Section \ref{sec:relativistic} do however allow us to postulate that this should be the case for supersymmetric, integrable theories. It is also natural to expect, by analogy with four dimensions, that this should also be true for theories that are just supersymmetric -- however, we have not analyzed any models in this class. Furthermore, we found evidence that cut-constructibility also partially works for integrable field theories without supersymmetry. For these models the missing rational terms are proportional to the tree-level S-matrix and therefore can be understood as a finite shift in the coupling.

The cut-constructible piece of the one-loop world-sheet S-matrix for the light-cone gauge-fixed superstring on AdS$_5 \times S^5$, which we computed in Section \ref{sec:strings}, matches perfectly with the result following from integrability. 
It is therefore hopeful that this method would work for other integrable string backgrounds \cite{Zarembo:2010sg,Cagnazzo:2012se}, for example AdS$_2 \times S^2 \times T^6$, AdS$_3 \times S^3 \times T^4$, AdS$_3\times S^3 \times S^3 \times S^1$  and AdS$_4 \times \mathbb{C}\mathbf{P}^3$ -- expressions for some tree-level and one-loop amplitudes for these theories are contained in \cite{Zarembo:2009au,Kalousios:2009ey,Rughoonauth:2012qd,Sundin:2013ypa,Hoare:2013pma}.  It would also be interesting to apply analogous unitarity techniques to other physical world-sheet observables, for example, in the form factor program initiated in \cite{Klose:2012ju}.

Finally, the natural extension of this work would be to generalize to both higher loops and higher points. The latter would be of particular interest in the case of non-integrable theories, and would necessarily involve a deeper understanding of rational terms.

\textbf{Note added:} We refer the reader to the related paper \cite{Engelund:2013fja}. In this work the authors independently proposed the idea of and developed (to two loops) generalized unitarity techniques applied to two-dimensional S-matrices. These techniques were used to compute the logarithmic terms of the one- and two-loop four-particle world-sheet S-matrix for the massive sectors of string theory on AdS$_3 \times S^3 \times T^4$, AdS$_3 \times S^3 \times S^3 \times S^1$, AdS$_4 \times \mathbb{C}\mathbf{P}^3$  and AdS$_5 \times S^5$ finding agreement with previous conjectures and results. At one-loop the two derivations are similar, however the contribution of the $t$-channel cut (amounting to rational terms) is fixed in a  different way -- in \cite{Engelund:2013fja} it is fixed by symmetries, whereas here we use a prescription following from the cutting procedure.


\section*{Acknowledgments}
\setcounter {equation} {0}

We are grateful to Valentina G. M. Puletti, Johannes Henn, Thomas Klose, Tristan McLoughlin, Marco Meineri, Jan Plefka, Radu Roiban, Domenico Seminara and Roberto Tateo for useful discussions.
We particularly thank Radu Roiban and Arkady Tseytlin for valuable comments on the draft. 
This work is funded by DFG via the Emmy Noether Program ``Gauge Fields from Strings''. 


\bibliographystyle{nb}

\bibliography{ref_Cuts2d}

\begin{thebibliography}{10}
\ifx\href\asklfhas\newcommand{\href}[2]{#2}\fi
\ifx\arxivref\asklfhas\newcommand{\arxivref}[2]{\href{http://arxiv.org/abs/#1}{#2}}\fi
\ifx\doiref\asklfhas\newcommand{\doiref}[2]{\href{http://dx.doi.org/#1}{#2}}\fi
\raggedright
\small
\parskip 0pt

\bibitem{Roib_Review}
\textit{``{Scattering amplitudes in gauge theories: progress and outlook}''},
edited by R.~Roiban, M.~Spradlin and A.~Volovich,
\textsf{\doiref{doi:10.1088/1751-8113/44/45/450301}{J.Phys.~A44,~450301~(2011)}}.

\bibitem{Beisert:2005tm}
N.~Beisert,
\textit{``{The SU$(2|2)$ dynamic S-matrix}''},
\textsf{Adv.Theor.Math.Phys.~12,~945~(2008)},
\texttt{\arxivref{hep-th/0511082}{hep-th/0511082}}.

\bibitem{Arutyunov:2009ga}
G.~Arutyunov and S.~Frolov,
\textit{``{Foundations of the AdS$_5 \times S^5$ Superstring. Part I}''},
\textsf{\doiref{10.1088/1751-8113/42/25/254003}{J.Phys.~A42,~254003~(2009)}},
\texttt{\arxivref{0901.4937}{arXiv:0901.4937}}.

\bibitem{Beisert:2010jr}
N.~Beisert, C.~Ahn, L.~F.~Alday, Z.~Bajnok, J.~M.~Drummond et~al.,
\textit{``{Review of AdS/CFT Integrability: An Overview}''},
\textsf{\doiref{10.1007/s11005-011-0529-2}{Lett.Math.Phys.~99,~3~(2012)}},
\texttt{\arxivref{1012.3982}{arXiv:1012.3982}}.

\bibitem{Klose:2006dd}
T.~Klose and K.~Zarembo,
\textit{``{Bethe ansatz in stringy sigma models}''},
\textsf{\doiref{10.1088/1742-5468/2006/05/P05006}{J.Stat.Mech.~0605,~P05006~(2006)}},
\texttt{\arxivref{hep-th/0603039}{hep-th/0603039}}.

\bibitem{Roiban:2006yc}
R.~Roiban, A.~Tirziu and A.~A.~Tseytlin,
\textit{``{Asymptotic Bethe ansatz S-matrix and Landau-Lifshitz type effective
  2-d actions}''},
\textsf{\doiref{10.1088/0305-4470/39/41/S19}{J.Phys.~A39,~13129~(2006)}},
\texttt{\arxivref{hep-th/0604199}{hep-th/0604199}}.

\bibitem{Klose:2006zd}
T.~Klose, T.~McLoughlin, R.~Roiban and K.~Zarembo,
\textit{``{Worldsheet scattering in AdS$_5 \times S^5$}''},
\textsf{\doiref{10.1088/1126-6708/2007/03/094}{JHEP~0703,~094~(2007)}},
\texttt{\arxivref{hep-th/0611169}{hep-th/0611169}}.

\bibitem{Klose:2007wq}
T.~Klose and K.~Zarembo,
\textit{``{Reduced sigma-model on AdS$_5 \times S^5$: One-loop scattering
  amplitudes}''},
\textsf{\doiref{10.1088/1126-6708/2007/02/071}{JHEP~0702,~071~(2007)}},
\texttt{\arxivref{hep-th/0701240}{hep-th/0701240}}.

\bibitem{Klose:2007rz}
T.~Klose, T.~McLoughlin, J.~Minahan and K.~Zarembo,
\textit{``{World-sheet scattering in AdS$_5 \times S^5$ at two loops}''},
\textsf{\doiref{10.1088/1126-6708/2007/08/051}{JHEP~0708,~051~(2007)}},
\texttt{\arxivref{0704.3891}{arXiv:0704.3891}}.

\bibitem{Zamolodchikov:1978xm}
A.~B.~Zamolodchikov and A.~B.~Zamolodchikov,
\textit{``{Factorized S-Matrices in Two-Dimensions as the Exact Solutions of
  Certain Relativistic Quantum Field Models}''},
\textsf{\doiref{10.1016/0003-4916(79)90391-9}{Annals~Phys.~120,~253~(1979)}}.

\bibitem{Ogievetsky:1987vv}
E.~Ogievetsky, P.~Wiegmann and N.~Reshetikhin,
\textit{``{The Principal Chiral Field in Two-Dimensions on Classical Lie
  Algebras: The Bethe Ansatz Solution and Factorized Theory of Scattering}''},
\textsf{\doiref{10.1016/0550-3213(87)90138-6}{Nucl.Phys.~B280,~45~(1987)}}.

\bibitem{Dorey:1996gd}
P.~Dorey,
\textit{``{Exact S matrices}''},
\texttt{\arxivref{hep-th/9810026}{hep-th/9810026}}.

\bibitem{Arefeva:1974bk}
I.~Arefeva and V.~Korepin,
\textit{``{Scattering in two-dimensional model with Lagrangian $(1/\gamma)
  ((\partial_\mu u)^2/2 + m^2 \cos(u-1))$}''},
\textsf{Pisma~Zh.Eksp.Teor.Fiz.~20,~680~(1974)}.

\bibitem{Britto:2004nc}
R.~Britto, F.~Cachazo and B.~Feng,
\textit{``{Generalized unitarity and one-loop amplitudes in $\mathcal N=4$
  super-Yang-Mills}''},
\textsf{\doiref{10.1016/j.nuclphysb.2005.07.014}{Nucl.Phys.~B725,~275~(2005)}},
\texttt{\arxivref{hep-th/0412103}{hep-th/0412103}}.

\bibitem{Bern:2011qt}
Z.~Bern and Y.-t.~Huang,
\textit{``{Basics of Generalized Unitarity}''},
\textsf{\doiref{10.1088/1751-8113/44/45/454003}{J.Phys.~A44,~454003~(2011)}},
\texttt{\arxivref{1103.1869}{arXiv:1103.1869}}.

\bibitem{vanNeerven:1985xr}
W.~van~Neerven,
\textit{``{Dimensional regularization of mass and infrared singularities in two
  loop on-shell vertex functions}''},
\textsf{\doiref{10.1016/0550-3213(86)90165-3}{Nucl.Phys.~B268,~453~(1986)}}.

\bibitem{Bern:1994zx}
Z.~Bern, L.~J.~Dixon, D.~C.~Dunbar and D.~A.~Kosower,
\textit{``{One loop n point gauge theory amplitudes, unitarity and collinear
  limits}''},
\textsf{\doiref{10.1016/0550-3213(94)90179-1}{Nucl.Phys.~B425,~217~(1994)}},
\texttt{\arxivref{hep-ph/9403226}{hep-ph/9403226}}.

\bibitem{Bern:1994cg}
Z.~Bern, L.~J.~Dixon, D.~C.~Dunbar and D.~A.~Kosower,
\textit{``{Fusing gauge theory tree amplitudes into loop amplitudes}''},
\textsf{\doiref{10.1016/0550-3213(94)00488-Z}{Nucl.Phys.~B435,~59~(1995)}},
\texttt{\arxivref{hep-ph/9409265}{hep-ph/9409265}}.

\bibitem{Dixon:1996wi}
L.~J.~Dixon,
\textit{``{Calculating scattering amplitudes efficiently}''},
\texttt{\arxivref{hep-ph/9601359}{hep-ph/9601359}}.

\bibitem{Eden:1966}
R.~J.~Eden, P.~V.~Landshoff, D.~I.~Olive and J.~C.~Polkinghorne,
\textit{``{The analytic S-matrix}''}.

\bibitem{Bern:1997nh}
Z.~Bern, J.~Rozowsky and B.~Yan,
\textit{``{Two loop four gluon amplitudes in $\mathcal N=4$
  superYang-Mills}''},
\textsf{\doiref{10.1016/S0370-2693(97)00413-9}{Phys.Lett.~B401,~273~(1997)}},
\texttt{\arxivref{hep-ph/9702424}{hep-ph/9702424}}.

\bibitem{Bern:1995db}
Z.~Bern and A.~Morgan,
\textit{``{Massive loop amplitudes from unitarity}''},
\textsf{\doiref{10.1016/0550-3213(96)00078-8}{Nucl.Phys.~B467,~479~(1996)}},
\texttt{\arxivref{hep-ph/9511336}{hep-ph/9511336}}.

\bibitem{Hollowood:1994vx}
T.~J.~Hollowood, J.~L.~Miramontes and Q.-H.~Park,
\textit{``{Massive integrable soliton theories}''},
\textsf{\doiref{10.1016/0550-3213(95)00142-F}{Nucl.Phys.~B445,~451~(1995)}},
\texttt{\arxivref{hep-th/9412062}{hep-th/9412062}}.

\bibitem{Bakas:1995bm}
I.~Bakas, Q.-H.~Park and H.-J.~Shin,
\textit{``{Lagrangian formulation of symmetric space sine-Gordon models}''},
\textsf{\doiref{10.1016/0370-2693(96)00026-3}{Phys.Lett.~B372,~45~(1996)}},
\texttt{\arxivref{hep-th/9512030}{hep-th/9512030}}.

\bibitem{Dorey:1994mg}
N.~Dorey and T.~J.~Hollowood,
\textit{``{Quantum scattering of charged solitons in the complex sine-Gordon
  model}''},
\textsf{\doiref{10.1016/0550-3213(95)00074-3}{Nucl.Phys.~B440,~215~(1995)}},
\texttt{\arxivref{hep-th/9410140}{hep-th/9410140}}.

\bibitem{Hoare:2010fb}
B.~Hoare and A.~Tseytlin,
\textit{``{On the perturbative S-matrix of generalized sine-Gordon models}''},
\textsf{\doiref{10.1007/JHEP11(2010)111}{JHEP~1011,~111~(2010)}},
\texttt{\arxivref{1008.4914}{arXiv:1008.4914}}.

\bibitem{deVega:1981ka}
H.~de~Vega and J.~Maillet,
\textit{``{Renormalization character and quantum S-matrix for a classically
  integrable theory}''},
\textsf{\doiref{10.1016/0370-2693(81)90049-6}{Phys.Lett.~B101,~302~(1981)}}.

\bibitem{Hollowood:2010dt}
T.~J.~Hollowood and J.~L.~Miramontes,
\textit{``{Classical and Quantum Solitons in the Symmetric Space Sine-Gordon
  Theories}''},
\textsf{\doiref{10.1007/JHEP04(2011)119}{JHEP~1104,~119~(2011)}},
\texttt{\arxivref{1012.0716}{arXiv:1012.0716}}.

\bibitem{Felder:1994be}
G.~Felder,
\textit{``{Elliptic quantum groups}''},
\texttt{\arxivref{hep-th/9412207}{hep-th/9412207}}.

\bibitem{Hoare:2013ysa}
B.~Hoare, T.~J.~Hollowood and J.~L.~Miramontes,
\textit{``{Restoring Unitarity in the q-Deformed World-Sheet S-Matrix}''},
\texttt{\arxivref{1303.1447}{arXiv:1303.1447}}.

\bibitem{Witten:1983ar}
E.~Witten,
\textit{``{Nonabelian Bosonization in Two-Dimensions}''},
\textsf{\doiref{10.1007/BF01215276}{Commun.Math.Phys.~92,~455~(1984)}}.

\bibitem{Knizhnik:1984nr}
V.~Knizhnik and A.~Zamolodchikov,
\textit{``{Current Algebra and Wess-Zumino Model in Two-Dimensions}''},
\textsf{\doiref{10.1016/0550-3213(84)90374-2}{Nucl.Phys.~B247,~83~(1984)}}.

\bibitem{Leutwyler:1991tv}
H.~Leutwyler and M.~A.~Shifman,
\textit{``{Perturbation theory in the Wess-Zumino-Novikov-Witten model}''},
\textsf{\doiref{10.1142/S0217751X92000387}{Int.J.Mod.Phys.~A7,~795~(1992)}}.

\bibitem{Tseytlin:1992ri}
A.~A.~Tseytlin,
\textit{``{Effective action of gauged WZW model and exact string solutions}''},
\textsf{\doiref{10.1016/0550-3213(93)90511-M}{Nucl.Phys.~B399,~601~(1993)}},
\texttt{\arxivref{hep-th/9301015}{hep-th/9301015}}.

\bibitem{Tseytlin:1993my}
A.~A.~Tseytlin,
\textit{``{Conformal sigma models corresponding to gauged Wess-Zumino-Witten
  theories}''},
\textsf{\doiref{10.1016/0550-3213(94)90461-8}{Nucl.Phys.~B411,~509~(1994)}},
\texttt{\arxivref{hep-th/9302083}{hep-th/9302083}}.

\bibitem{deWit:1993qv}
B.~de~Wit, M.~T.~Grisaru and P.~van~Nieuwenhuizen,
\textit{``{The WZNW model at two loops}''},
\textsf{\doiref{10.1016/0550-3213(93)90537-Y}{Nucl.Phys.~B408,~299~(1993)}},
\texttt{\arxivref{hep-th/9307027}{hep-th/9307027}}.

\bibitem{Pohlmeyer:1975nb}
K.~Pohlmeyer,
\textit{``{Integrable Hamiltonian Systems and Interactions Through Quadratic
  Constraints}''},
\textsf{\doiref{10.1007/BF01609119}{Commun.Math.Phys.~46,~207~(1976)}}.

\bibitem{Eichenherr:1979yw}
H.~Eichenherr and K.~Pohlmeyer,
\textit{``{Lax pairs for certain generalizations of the sine-Gordon
  equation}''},
\textsf{\doiref{10.1016/0370-2693(79)90079-0}{Phys.Lett.~B89,~76~(1979)}}.

\bibitem{Tseytlin:2003ii}
A.~A.~Tseytlin,
\textit{``{Spinning strings and AdS/CFT duality}''},
\texttt{\arxivref{hep-th/0311139}{hep-th/0311139}}.

\bibitem{Barbashov:1980kz}
B.~Barbashov and V.~Nesterenko,
\textit{``{Relativistic string model in a space-time of a constant
  curvature}''},
\textsf{\doiref{10.1007/BF02046761}{Commun.Math.Phys.~78,~499~(1981)}}.

\bibitem{DeVega:1992xc}
H.~De~Vega and N.~G.~Sanchez,
\textit{``{Exact integrability of strings in D-Dimensional De Sitter
  space-time}''},
\textsf{\doiref{10.1103/PhysRevD.47.3394}{Phys.Rev.~D47,~3394~(1993)}}.

\bibitem{Metsaev:1998it}
R.~Metsaev and A.~A.~Tseytlin,
\textit{``{Type IIB superstring action in AdS$_5 \times S^5$ background}''},
\textsf{\doiref{10.1016/S0550-3213(98)00570-7}{Nucl.Phys.~B533,~109~(1998)}},
\texttt{\arxivref{hep-th/9805028}{hep-th/9805028}}.

\bibitem{Grigoriev:2007bu}
M.~Grigoriev and A.~A.~Tseytlin,
\textit{``{Pohlmeyer reduction of AdS$_5 \times S^5$ superstring sigma
  model}''},
\textsf{\doiref{10.1016/j.nuclphysb.2008.01.006}{Nucl.Phys.~B800,~450~(2008)}},
\texttt{\arxivref{0711.0155}{arXiv:0711.0155}}.

\bibitem{Mikhailov:2007xr}
A.~Mikhailov and S.~Schafer-Nameki,
\textit{``{Sine-Gordon-like action for the Superstring in AdS$_5 \times
  S^5$}''},
\textsf{\doiref{10.1088/1126-6708/2008/05/075}{JHEP~0805,~075~(2008)}},
\texttt{\arxivref{0711.0195}{arXiv:0711.0195}}.

\bibitem{Grigoriev:2008jq}
M.~Grigoriev and A.~A.~Tseytlin,
\textit{``{On reduced models for superstrings on AdS$_n \times S^n$}''},
\textsf{\doiref{10.1142/S0217751X08040652}{Int.J.Mod.Phys.~A23,~2107~(2008)}},
\texttt{\arxivref{0806.2623}{arXiv:0806.2623}}.

\bibitem{Babichenko:2009dk}
A.~Babichenko, J.~Stefanski,~B. and K.~Zarembo,
\textit{``{Integrability and the AdS$_3$/CFT$_2$ correspondence}''},
\textsf{\doiref{10.1007/JHEP03(2010)058}{JHEP~1003,~058~(2010)}},
\texttt{\arxivref{0912.1723}{arXiv:0912.1723}}.

\bibitem{Sorokin:2011rr}
D.~Sorokin, A.~Tseytlin, L.~Wulff and K.~Zarembo,
\textit{``{Superstrings in AdS$_2 \times S^2 \times T^6$}''},
\textsf{\doiref{10.1088/1751-8113/44/27/275401}{J.Phys.~A44,~275401~(2011)}},
\texttt{\arxivref{1104.1793}{arXiv:1104.1793}}.

\bibitem{Roiban:2009vh}
R.~Roiban and A.~A.~Tseytlin,
\textit{``{UV finiteness of Pohlmeyer-reduced form of the AdS$_5 \times S^5$
  superstring theory}''},
\textsf{\doiref{10.1088/1126-6708/2009/04/078}{JHEP~0904,~078~(2009)}},
\texttt{\arxivref{0902.2489}{arXiv:0902.2489}}.

\bibitem{Schmidtt:2010bi}
D.~M.~Schmidtt,
\textit{``{Supersymmetry Flows, Semi-Symmetric Space Sine-Gordon Models And The
  Pohlmeyer Reduction}''},
\textsf{\doiref{10.1007/JHEP03(2011)021}{JHEP~1103,~021~(2011)}},
\texttt{\arxivref{1012.4713}{arXiv:1012.4713}}.

\bibitem{Hollowood:2011fq}
T.~J.~Hollowood and J.~L.~Miramontes,
\textit{``{The AdS$_5 \times S_5$ Semi-Symmetric Space Sine-Gordon Theory}''},
\textsf{\doiref{10.1007/JHEP05(2011)136}{JHEP~1105,~136~(2011)}},
\texttt{\arxivref{1104.2429}{arXiv:1104.2429}}.

\bibitem{Goykhman:2011mq}
M.~Goykhman and E.~Ivanov,
\textit{``{Worldsheet Supersymmetry of Pohlmeyer-Reduced AdS$_n \times S^n$
  Superstrings}''},
\textsf{\doiref{10.1007/JHEP09(2011)078}{JHEP~1109,~078~(2011)}},
\texttt{\arxivref{1104.0706}{arXiv:1104.0706}}.

\bibitem{Schmidtt:2011nr}
D.~M.~Schmidtt,
\textit{``{Integrability vs Supersymmetry: Poisson Structures of The Pohlmeyer
  Reduction}''},
\textsf{\doiref{10.1007/JHEP11(2011)067}{JHEP~1111,~067~(2011)}},
\texttt{\arxivref{1106.4796}{arXiv:1106.4796}}.

\bibitem{Hoare:2009fs}
B.~Hoare and A.~Tseytlin,
\textit{``{Tree-level S-matrix of Pohlmeyer reduced form of AdS$_5 \times S^5$
  superstring theory}''},
\textsf{\doiref{10.1007/JHEP02(2010)094}{JHEP~1002,~094~(2010)}},
\texttt{\arxivref{0912.2958}{arXiv:0912.2958}}.

\bibitem{Hoare:2011fj}
B.~Hoare and A.~Tseytlin,
\textit{``{Towards the quantum S-matrix of the Pohlmeyer reduced version of
  AdS$_5 \times S^5$ superstring theory}''},
\textsf{\doiref{10.1016/j.nuclphysb.2011.05.016}{Nucl.Phys.~B851,~161~(2011)}},
\texttt{\arxivref{1104.2423}{arXiv:1104.2423}}.

\bibitem{Kobayashi:1991rh}
K.-i.~Kobayashi and T.~Uematsu,
\textit{``{S matrix of N=2 supersymmetric Sine-Gordon theory}''},
\textsf{\doiref{10.1016/0370-2693(92)91603-7}{Phys.Lett.~B275,~361~(1992)}},
\texttt{\arxivref{hep-th/9110040}{hep-th/9110040}}.

\bibitem{Shankar:1977cm}
R.~Shankar and E.~Witten,
\textit{``{The S Matrix of the Supersymmetric Nonlinear Sigma Model}''},
\textsf{\doiref{10.1103/PhysRevD.17.2134}{Phys.Rev.~D17,~2134~(1978)}}.

\bibitem{Bena:2003wd}
I.~Bena, J.~Polchinski and R.~Roiban,
\textit{``{Hidden symmetries of the AdS$_5 \times S^5$ superstring}''},
\textsf{\doiref{10.1103/PhysRevD.69.046002}{Phys.Rev.~D69,~046002~(2004)}},
\texttt{\arxivref{hep-th/0305116}{hep-th/0305116}}.

\bibitem{Kazakov:2004qf}
V.~Kazakov, A.~Marshakov, J.~Minahan and K.~Zarembo,
\textit{``{Classical/quantum integrability in AdS/CFT}''},
\textsf{\doiref{10.1088/1126-6708/2004/05/024}{JHEP~0405,~024~(2004)}},
\texttt{\arxivref{hep-th/0402207}{hep-th/0402207}}.

\bibitem{Berkovits:2004xu}
N.~Berkovits,
\textit{``{Quantum consistency of the superstring in AdS$_5 \times S^5$
  background}''},
\textsf{\doiref{10.1088/1126-6708/2005/03/041}{JHEP~0503,~041~(2005)}},
\texttt{\arxivref{hep-th/0411170}{hep-th/0411170}}.

\bibitem{Arutyunov:2004vx}
G.~Arutyunov, S.~Frolov and M.~Staudacher,
\textit{``{Bethe ansatz for quantum strings}''},
\textsf{\doiref{10.1088/1126-6708/2004/10/016}{JHEP~0410,~016~(2004)}},
\texttt{\arxivref{hep-th/0406256}{hep-th/0406256}}.

\bibitem{Janik:2006dc}
R.~A.~Janik,
\textit{``{The AdS$_5 \times S^5$ superstring worldsheet S-matrix and crossing
  symmetry}''},
\textsf{\doiref{10.1103/PhysRevD.73.086006}{Phys.Rev.~D73,~086006~(2006)}},
\texttt{\arxivref{hep-th/0603038}{hep-th/0603038}}.

\bibitem{Volin:2009uv}
D.~Volin,
\textit{``{Minimal solution of the AdS/CFT crossing equation}''},
\textsf{\doiref{10.1088/1751-8113/42/37/372001}{J.Phys.~A42,~372001~(2009)}},
\texttt{\arxivref{0904.4929}{arXiv:0904.4929}}.

\bibitem{Beisert:2006ib}
N.~Beisert, R.~Hernandez and E.~Lopez,
\textit{``{A Crossing-symmetric phase for AdS$_5 \times S^5$ strings}''},
\textsf{\doiref{10.1088/1126-6708/2006/11/070}{JHEP~0611,~070~(2006)}},
\texttt{\arxivref{hep-th/0609044}{hep-th/0609044}}.

\bibitem{Beisert:2006ez}
N.~Beisert, B.~Eden and M.~Staudacher,
\textit{``{Transcendentality and Crossing}''},
\textsf{\doiref{10.1088/1742-5468/2007/01/P01021}{J.Stat.Mech.~0701,~P01021~(2007)}},
\texttt{\arxivref{hep-th/0610251}{hep-th/0610251}}.

\bibitem{McLoughlin:2010jw}
T.~McLoughlin,
\textit{``{Review of AdS/CFT Integrability, Chapter II.2: Quantum Strings in
  AdS$_5 \times S^5$}''},
\textsf{\doiref{10.1007/s11005-011-0510-0}{Lett.Math.Phys.~99,~127~(2012)}},
\texttt{\arxivref{1012.3987}{arXiv:1012.3987}}.

\bibitem{Arutyunov:2006ak}
G.~Arutyunov, S.~Frolov, J.~Plefka and M.~Zamaklar,
\textit{``{The Off-shell Symmetry Algebra of the Light-cone AdS$_5 \times S^5$
  Superstring}''},
\textsf{\doiref{10.1088/1751-8113/40/13/018}{J.Phys.~A40,~3583~(2007)}},
\texttt{\arxivref{hep-th/0609157}{hep-th/0609157}}.

\bibitem{Arutyunov:2006gs}
G.~Arutyunov, S.~Frolov and M.~Zamaklar,
\textit{``{Finite-size Effects from Giant Magnons}''},
\textsf{\doiref{10.1016/j.nuclphysb.2006.12.026}{Nucl.Phys.~B778,~1~(2007)}},
\texttt{\arxivref{hep-th/0606126}{hep-th/0606126}}.

\bibitem{Frolov:2006cc}
S.~Frolov, J.~Plefka and M.~Zamaklar,
\textit{``{The AdS$_5 \times S^5$ superstring in light-cone gauge and its Bethe
  equations}''},
\textsf{\doiref{10.1088/0305-4470/39/41/S15}{J.Phys.~A39,~13037~(2006)}},
\texttt{\arxivref{hep-th/0603008}{hep-th/0603008}}.

\bibitem{Roiban:2007jf}
R.~Roiban, A.~Tirziu and A.~A.~Tseytlin,
\textit{``{Two-loop world-sheet corrections in $AdS_5 \times S^5$
  superstring}''},
\textsf{\doiref{10.1088/1126-6708/2007/07/056}{JHEP~0707,~056~(2007)}},
\texttt{\arxivref{0704.3638}{arXiv:0704.3638}}.

\bibitem{Maldacena:2006rv}
J.~M.~Maldacena and I.~Swanson,
\textit{``{Connecting giant magnons to the pp-wave: An Interpolating limit of
  AdS$_5 \times S^5$}''},
\textsf{\doiref{10.1103/PhysRevD.76.026002}{Phys.Rev.~D76,~026002~(2007)}},
\texttt{\arxivref{hep-th/0612079}{hep-th/0612079}}.

\bibitem{Puletti:2007hq}
V.~Giangreco Marotta~Puletti, T.~Klose and O.~Ohlsson~Sax,
\textit{``{Factorized world-sheet scattering in near-flat AdS$_5 \times
  S^5$}''},
\textsf{\doiref{10.1016/j.nuclphysb.2007.09.018}{Nucl.Phys.~B792,~228~(2008)}},
\texttt{\arxivref{0707.2082}{arXiv:0707.2082}}.

\bibitem{Arutyunov:2006yd}
G.~Arutyunov, S.~Frolov and M.~Zamaklar,
\textit{``{The Zamolodchikov-Faddeev algebra for AdS$_5 \times S^5$
  superstring}''},
\textsf{\doiref{10.1088/1126-6708/2007/04/002}{JHEP~0704,~002~(2007)}},
\texttt{\arxivref{hep-th/0612229}{hep-th/0612229}}.

\bibitem{Beisert:2004hm}
N.~Beisert, V.~Dippel and M.~Staudacher,
\textit{``{A Novel long range spin chain and planar $\mathcal N=4$ super
  Yang-Mills}''},
\textsf{\doiref{10.1088/1126-6708/2004/07/075}{JHEP~0407,~075~(2004)}},
\texttt{\arxivref{hep-th/0405001}{hep-th/0405001}}.

\bibitem{Arutyunov:2006iu}
G.~Arutyunov and S.~Frolov,
\textit{``{On AdS$_5 \times S^5$ String S-matrix}''},
\textsf{\doiref{10.1016/j.physletb.2006.06.064}{Phys.Lett.~B639,~378~(2006)}},
\texttt{\arxivref{hep-th/0604043}{hep-th/0604043}}.

\bibitem{Ahn:2010ka}
C.~Ahn and R.~I.~Nepomechie,
\textit{``{Review of AdS/CFT Integrability, Chapter III.2: Exact World-Sheet
  S-Matrix}''},
\textsf{\doiref{10.1007/s11005-011-0478-9}{Lett.Math.Phys.~99,~209~(2012)}},
\texttt{\arxivref{1012.3991}{arXiv:1012.3991}}.

\bibitem{Zarembo:2010sg}
K.~Zarembo,
\textit{``{Strings on Semisymmetric Superspaces}''},
\textsf{\doiref{10.1007/JHEP05(2010)002}{JHEP~1005,~002~(2010)}},
\texttt{\arxivref{1003.0465}{arXiv:1003.0465}}.

\bibitem{Cagnazzo:2012se}
A.~Cagnazzo and K.~Zarembo,
\textit{``{B-field in AdS$_3$/CFT$_2$ Correspondence and Integrability}''},
\textsf{\doiref{10.1007/JHEP11(2012)133}{JHEP~1211,~133~(2012)}},
\texttt{\arxivref{1209.4049}{arXiv:1209.4049}}.

\bibitem{Zarembo:2009au}
K.~Zarembo,
\textit{``{Worldsheet spectrum in AdS$_4$/CFT$_3$ correspondence}''},
\textsf{\doiref{10.1088/1126-6708/2009/04/135}{JHEP~0904,~135~(2009)}},
\texttt{\arxivref{0903.1747}{arXiv:0903.1747}}.

\bibitem{Kalousios:2009ey}
C.~Kalousios, C.~Vergu and A.~Volovich,
\textit{``{Factorized Tree-level Scattering in AdS$_4 \times
  \mathbb{C}\mathbf{P}^3$}''},
\textsf{\doiref{10.1088/1126-6708/2009/09/049}{JHEP~0909,~049~(2009)}},
\texttt{\arxivref{0905.4702}{arXiv:0905.4702}}.

\bibitem{Rughoonauth:2012qd}
N.~Rughoonauth, P.~Sundin and L.~Wulff,
\textit{``{Near BMN dynamics of the AdS$_3 \times S^3 \times S^3 \times S^1$
  superstring}''},
\textsf{\doiref{10.1007/JHEP07(2012)159}{JHEP~1207,~159~(2012)}},
\texttt{\arxivref{1204.4742}{arXiv:1204.4742}}.

\bibitem{Sundin:2013ypa}
P.~Sundin and L.~Wulff,
\textit{``{Worldsheet scattering in AdS$_3$/CFT$_2$}''},
\texttt{\arxivref{1302.5349}{arXiv:1302.5349}}.

\bibitem{Hoare:2013pma}
B.~Hoare and A.~Tseytlin,
\textit{``{On string theory on AdS$_3 \times S^3 \times T^4$ with mixed 3-form
  flux: tree-level S-matrix}''},
\texttt{\arxivref{1303.1037}{arXiv:1303.1037}}.

\bibitem{Klose:2012ju}
T.~Klose and T.~McLoughlin,
\textit{``{Worldsheet Form Factors in AdS/CFT}''},
\textsf{\doiref{10.1103/PhysRevD.87.026004}{Phys.Rev.~D87,~026004~(2013)}},
\texttt{\arxivref{1208.2020}{arXiv:1208.2020}}.

\bibitem{Engelund:2013fja}
O.~T.~Engelund, R.~W.~McKeown and R.~Roiban,
\textit{``{Generalized unitarity and the worldsheet S matrix in $AdS_n \times
  S^n \times M^{10-2n}$}''},
\texttt{\arxivref{1304.4281}{arXiv:1304.4281}}.

\end{thebibliography}

\end{document}